\documentclass[%
 reprint,
bibnotes,
 amsmath,amssymb,
 aps,
prx,
]{revtex4-2}

\usepackage[normalem]{ulem}

\usepackage{graphicx}
\usepackage{dcolumn}
\usepackage{bm}
\usepackage[breaklinks,colorlinks,bookmarks=false,citecolor=blue,linkcolor=red,urlcolor=blue]{hyperref}
\usepackage{cleveref}
\usepackage[mathlines]{lineno}
\usepackage{algorithm}
\usepackage[noend]{algpseudocode}
\usepackage{xcolor}
\usepackage{pifont}
\usepackage{tikz}
\usetikzlibrary{matrix}

\usepackage{accents}

\usepackage{soul}

\newcommand{\chispin}{\chi_{\text{sp}}}
\newcommand{\chispinstag}{\chi_{\text{sp}}^{\text{st}}}
\newcommand{\chispinq}[1]{\chi_{\text{sp}}(\mathbf{#1})}
\newcommand{\chicharge}{\chi_{\text{ch}}}
\newcommand{\chichargeq}[1]{\chi_{\text{ch}}(\mathbf{#1})}
\newcommand{\structurespin}{S_{\text{sp}}}
\newcommand{\structurecharge}{S_{\text{ch}}}
\DeclareMathOperator*{\argmax}{argmax}

\begin{document}
\preprint{arXiv}

\title{The Weak, the Strong and the Long Correlation Regimes of the Two-Dimensional Hubbard Model at Finite Temperature}

\author{Fedor \v{S}imkovic IV$^{1,2}$}
\email{fsimkovic@gmail.com}
\author{Riccardo Rossi$^3$}
\author{Michel Ferrero$^{1,2}$}

\affiliation{
$^1$CPHT, CNRS, Ecole Polytechnique, Institut Polytechnique de Paris, Route de Saclay, 91128 Palaiseau, France\\
$^2$Coll\`ege de France, 11 place Marcelin Berthelot, 75005 Paris, France\\
$^3$Institute of Physics, Ecole Polytechnique F\'ed\'erale de Lausanne (EPFL), CH-1015 Lausanne, Switzerland}

\date{\today}

\begin{abstract}
We investigate the momentum-resolved spin and charge susceptibilities, as well as the chemical potential and double occupancy in the two-dimensional Hubbard model as functions of doping, temperature and interaction strength. Through these quantities, we identify a weak-coupling regime, a strong-coupling regime with short-range correlations and an intermediate-coupling regime with long magnetic correlation lengths.
In the spin channel, we observe an additional crossover from commensurate to incommensurate correlations.
In contrast, we find charge correlations to be only short ranged for all studied temperatures, which suggests that the spin and charge responses are decoupled. These findings were obtained by a novel connected determinant diagrammatic Monte Carlo algorithm for the computation of double expansions, which we introduce in this paper. This permits us to
obtain numerically exact results at unprecedentedly low temperatures $T\geq 0.067$ for interactions up to $U\leq 8$, while working on arbitrarily large lattices. Our method also allows us to gain physical insights from investigating the analytic structure of perturbative series. We connect to previous work by studying smaller lattice geometries and report substantial finite-size effects.
\end{abstract}

\maketitle

\section{Introduction} \label{intro}

The two-dimensional Fermi-Hubbard model~\cite{hubbard63, kanamori, anderson1963theory, qin2021hubbard, arovas2021hubbard} plays the role of the fruit fly within strongly-correlated electron models.
On the one hand, the model is a rich platform to investigate fundamental questions about
the properties of interacting quantum
systems. On the other hand, it is also
believed to be directly relevant to the study of correlated electronic materials.
It is for example conjectured to
capture essential physical properties of high-temperature superconductors such as cuprates~\cite{bednorz1986possible,orenstein2000advances} even though the exact connection is subject to ongoing debate~\cite{imadametal1998, nagaosadoping2006, keimer2015quantum}. Many relevant experiments find intricately convoluted orders of spin and charge stripes, charge and pair density waves, and unconventional superconductivity which undeniably adds to the difficulty of disentangling the roles of individual phenomena~\cite{dagotto2005complexity,aichhorn2007phase, fujita2002competition, abbamonte2005spatially, choubey2020atomic}.

While the Hubbard model has mainly been the subject of theoretical studies, it can now also be directly simulated at moderately-high temperatures by means of cold atoms trapped in optical lattices~\cite{jaksch1998cold, Bloch_review_2005, jordens2008mott, greif2015formation, cheuk2016observation, parsons2016site, greiner2017}. More recently, efforts in quantum computing technology have been aimed at the Hubbard model with the hope of finding effective algorithms for noisy near-term quantum hardware~\cite{arute2020observation, bauer2020quantum, cadestrategies2020, mcardle2020quantum, clinton2021hamiltonian, besserve2021unraveling}. Large collaborative projects have been formed with the aim of comparing the results of numerous state-of-the-art algorithms in order to better understand the model as well as provide unbiased consensus benchmarks for future experimental and theoretical studies~\cite{leblanc2015solutions, zheng2017stripe, qinabsence2020, M7}.
Despite the unquestionable and persistent interest of multiple scientific communities, a consensus phase diagram for the Hubbard model in the most intriguing parameter regimes is still missing~\cite{qin2021hubbard, arovas2021hubbard}.

Without doubt, a number of additional important unsolved questions remain to be answered before victory over the Hubbard model can be declared. To name but a few: Does the model host high-temperature superconductivity? If yes, how is it affected by competing magnetic instabilities? What is the relation between the onsets of strong spin and charge correlations? How does entering the strongly correlated regime manifest itself? Is a long correlation length a necessary ingredient?
What is the relationship between the pseudogap and the coupling strength or magnetic correlation
length?

Many previous works have addressed some of these questions.
One of the best understood regimes of the model is half-filling (one electron per lattice site on average), where the ground state is an antiferromagnetic insulator and the Mermin-Wagner theorem prohibits long-range order at any finite temperature. As a result, a series of crossovers form from a metallic regime to a quasi-antiferromagnetic insulator with exponentially large correlation lengths~\cite{schaeferfate2015, fedor_hf, kim_cdet, M7}, which becomes maximal at a particular finite interaction for any non-zero temperature. The nature of the insulating gap also changes as function of interaction, from Slater-like~\cite{slater1951magnetic} to Mott-like~\cite{mott1949basis}, and reaches the Heisenberg regime in the infinite-interaction limit~\cite{anderson1959new}.

Away from half-filling, the picture becomes a lot more blurry due to a much larger number of competing phases and the relative lack of controlled results from numerical algorithms. There are numerous methods used for extracting ground-state properties, such as Hartree-Fock theory (HF)~\cite{zaanencharged1989}, the density-matrix renormalization group  (DMRG)~\cite{whitestripes2003, huang2018stripe, jianground2020}, variational Monte Carlo (VMC)~\cite{idocompetition2018}, auxiliary field quantum Monte Carlo (AFQMC)~\cite{zheng2017stripe, sorella2021phase}, density matrix embedding theory (DMET)~\cite{knizia2012density, zheng2017stripe, zhengground2016} as well as inhomogeneous dynamical mean field theory (iDMFT)~\cite{peters_kawakami} and diagrammatic Monte Carlo (DiagMC)~\cite{deng}. The consensus picture is that spin and charge stripe-ordered phases~\cite{peters_kawakami, idocompetition2018, qinabsence2020, sorella2021phase}, superconducting phases~\cite{deng, simkovicsuperfluid2021,sorella2021phase}, as well as spatially phase separated states~\cite{macridin2006phase, aichhorn2007phase, chang2008spatially, galanakis2011quantum, sorella2015finite, idocompetition2018, sorella2021phase} can be found. It has been documented that superconducting phases seem predominant at weak and moderate coupling~\cite{deng, simkovicsuperfluid2021} while stripe-ordered phases win out at strong coupling~\cite{qinabsence2020, sorella2021phase}. Beyond those observations, a complete phase diagram is still lacking, mainly due to minuscule energy differences between competing states.

At finite temperature, one finds magnetic correlations in the spin and charge channels instead (together with superconducting ones). These have been documented by means of low-order diagrammatic Monte Carlo~\cite{simkovicmagnetic2017}, determinant Quantum Monte Carlo (DQMC)~\cite{huang2018stripe}, dynamical mean field theory (DMFT) and its cluster extensions ~\cite{fleck2000one,vilardi2018dynamically,vanhala2018dynamical, dash2021charge, maier_2021_fluctuating}, and by
minimally entangled typical thermal states (METTS)~\cite{wietekstripes2021}. These methods have also been able to capture the commensurate to incommensurate crossover, particularly in the spin channel, and as a function of temperature, doping and next-nearest-neighbor hopping. Despite these advancements, the relation between the appearance of strong correlations in the spin and charge channels is not yet fully understood. Another open question is the connection between the onset of the strongly correlated regime and the occurrence of long magnetic correlation lengths, which is particularly difficult to answer by simulating small finite-size clusters.

The biggest challenge in this regime of the Hubbard model is the conciliation of the finite-temperature results with the ground-state calculations~\cite{qin2021hubbard}. For many finite-temperature algorithms it is notoriously difficult to reach low enough temperatures to capture the relevant physics in order to make the connection, mainly due to the appearance of the fermionic sign problem. Almost all algorithms, whether ground state or finite-temperature, have to resolve to some sort of approximation. Notably, most of these cannot be formulated in the thermodynamic limit and thus suffer from finite-size effects~\cite{huang2018stripe, qinabsence2020, wietekstripes2021, sorella2021phase}, finite-momentum resolution~\cite{maier_2021_fluctuating}, and/or effects coming from the choice of boundary conditions. In particular, there is a number of methods restricted to the analysis of the Hubbard model on width-four cylinder geometries~\cite{huang2018stripe, qinabsence2020, wietekstripes2021}, whose correspondence to the model in the thermodynamic limit has not yet been thoroughly studied, mainly due to the lack of benchmarks from the latter.

\begin{figure}
\centering
\includegraphics[width=0.45\textwidth]{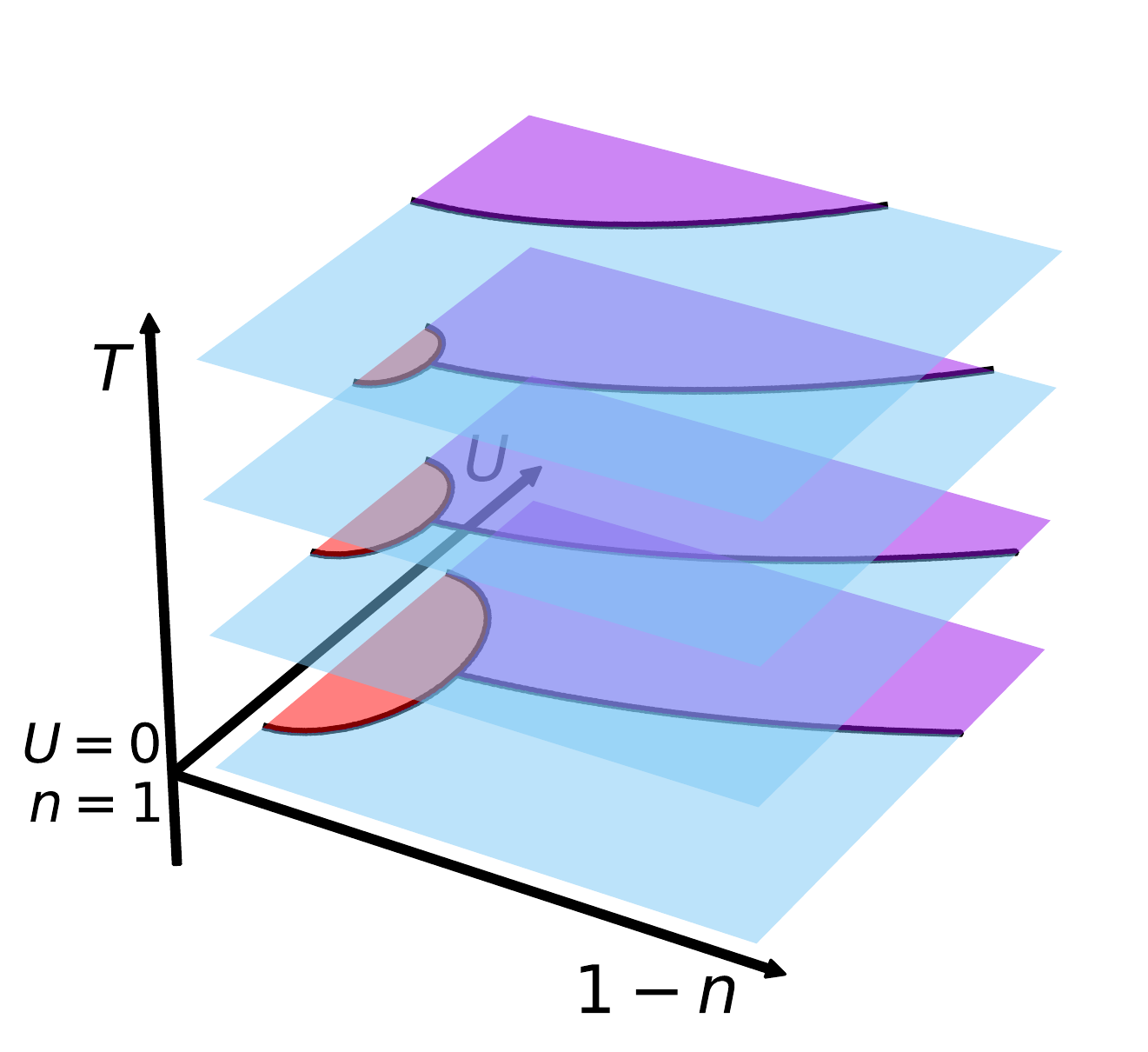}
\vspace{-1.25cm}

\raggedright
\includegraphics[width=0.20\textwidth]{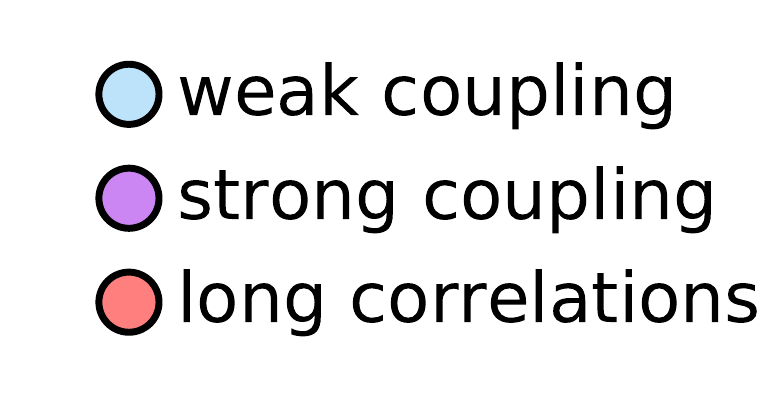}
\caption{Sketch of the crossovers between different regimes of correlations for the doped Hubbard model as a function of (non-zero) temperature $T$, doping $1-n$ and interaction strength $U$.}
\label{Fig:phase_diagram}
\end{figure}

In this work, we study the doped Hubbard model at finite temperature on large systems that do not suffer from finite-size effects This allows us to compute the spin and charge susceptibilities with a high momentum resolution, as functions of density, temperature and interaction strength.
We observe three distinct regimes as we move in parameter space, as sketched in Fig.~\ref{Fig:phase_diagram}. We denote the first regime \emph{weak
coupling} (blue). It is found predominantly at small interactions and has short correlation lengths.
Secondly, we observe a regime of what we call \emph{long correlations} (red) centered around half-filling and at finite interactions. The system is almost long-range ordered with its magnetic correlation length growing rapidly with decreasing temperature. Notably, we find the correlations to be most pronounced at significantly lower interaction strengths than the values usually cited by literature~\cite{cheuk2016observation, parsons2016site}. Finally, the \emph{strong
coupling} (purple) regime is found predominantly at higher values of interactions and in a broader range of dopings. The system is more localized, marked by low values of the double-occupancy and short-range correlations, with the correlation length being only mildly affected by decreasing temperature. We also report the appearance of a Mott gap at half-filling in this regime.

By evaluating the momentum-resolved spin susceptibility, we also investigate the nature of the magnetic ordering tendencies when the correlation length is sufficiently large, uncovering a commensurate to incommensurate crossover as a function of temperature, doping and interactions.
Additionally, we find that spin correlations have much higher onset temperatures as compared to charge correlations that never become long-ranged in the range of temperatures we studied.
This suggests that the spin and charge responses are decoupled over an extended region
of parameter space.

The findings presented in this manuscript were made possible by a novel connected determinant diagrammatic Monte Carlo algorithm (CDet) for the computation of perturbative double expansions, which, in particular, allows us to construct series at fixed density away from half-filling. This permits us to compute numerically-exact results for large lattice sizes and at unprecedentedly low finite temperatures down to $T \geq 0.067$ for interaction strengths $U \leq 8$. As our algorithm allows us to study arbitrary lattice sizes and geometries we are able to quantify the finite-size effects which are present in various other methods. Additionally, we show that much of the physics we have observed has a direct correspondence with the analytic properties of spin and charge susceptibilities as functions of {\it complex} interaction strength.

The paper is structured as follows: In Sec.~\ref{model} we introduce the Hubbard model and outline the main ingredients of our theoretical formalism. In Sec.~\ref{results} we present our key findings, focusing on mapping out the three distinct correlation regimes in Sec.~\ref{sec:stong_coupling}, the commensurate to incommensurate crossover in Sec.~\ref{sec:commensurate_to_incommensurate}, the decoupling of responses found in the spin and charge susceptibilities in Sec.~\ref{sec:charge}, we discuss finite size lattice effects in Sec.~\ref{sec:system_size}, and we elaborate physical insights from the analytic structure of perturbative series in Sec.~\ref{sec:poles}. We provide further details of our novel Diagrammatic Monte Carlo algorithm in Sec.~\ref{sec:methods} and finish with a cumulative discussion of our results together with an outlook for future studies in Sec.~\ref{conclusions}.

\section{Model and Formalism}\label{model}

\subsection{Hubbard model on the square lattice} \label{sec:hubbard}

The grand-canonical Hamiltonian for the Fermi-Hubbard model~\cite{hubbard63, kanamori, anderson1963theory, qin2021hubbard, arovas2021hubbard}  reads:
\begin{align}
\hat{H} = \sum_{\mathbf{k}, \sigma} \epsilon_{\mathbf{k}} \,
\hat{c}_{\mathbf{k} \sigma}^\dagger
\hat{c}_{\mathbf{k} \sigma}^{\phantom{\dagger}}+U\sum_{\mathbf{r}} \hat{n}_{\mathbf{r}\uparrow}\,
\hat{n}_{\mathbf{r} \downarrow}
-\mu \sum_{\mathbf{r},\sigma}\hat{n}_{\mathbf{r}\sigma},
\label{H}
\end{align}
where $\hat{c}^\dagger$ and $\hat{c}$ denote the fermionic creation and annihilation operators, $\mathbf{k}=(k_x,k_y)$ the reciprocal lattice momentum,
$\sigma \in \{ \uparrow, \downarrow \} $ the fermionic spin, $\mathbf{r} = (x,y)$ labels lattice sites, $U$ denotes the onsite
repulsion strength, $\mu$ the chemical potential, and the square lattice dispersion relation is given by $\epsilon_{\mathbf{k}}=-2\,t\left(\cos k_x+\cos k_y\right)$, where $t$ is the nearest-neighbor hopping amplitude (we set $t=1$ in our units), and $\hat{n}_{\mathbf{r}\sigma}$ counts the number of particles with spin $\sigma$ at site $\mathbf{r}$.

In the majority of this work (with the exception of Sec.~\ref{sec:system_size}) we will present results for the model on a periodic square lattice of fixed linear system size $L = 64$ as this is found to be sufficient to eliminate finite-size effects for the range of temperatures we consider. This allows to easily gather statistics for all the $L^2$ values needed to construct the momentum-resolved susceptibilities, but there exists no technical limitation for our algorithm in considering larger systems or even directly the thermodynamic limit. We study thermal equilibrium properties at non-zero temperature $T$. In the following, the thermal average of an operator $\hat{\mathcal{O}}$ is denoted by $\langle \hat{\mathcal{O}} \rangle=\tfrac{1}{Z} \text{Tr}\,e^{-\hat{H}/T}\hat{\mathcal{O}}$ where the grand-canonical partition function is defined as $Z =\text{Tr}\, e^{-\hat{H}/T}$, and we also use the imaginary-time Heisenberg picture for operators, $\hat{\mathcal{O}}(\tau)=e^{\tau\hat{H}}\,\hat{\mathcal{O}}\,e^{-\tau \hat{H}}$.

\subsection{Probes of spin and charge correlations}
The main probes of spin and charge correlations we use in this work are the momentum-resolved spin susceptibility (in real and reciprocal space),
which provides a quantitative measure of magnetic ordering tendencies, and the momentum-resolved charge susceptibility, which characterizes the response of the system to an inhomogeneous density perturbation.
The spin susceptibility in real space is:
\begin{align}
\chispin(\mathbf{r})
=\int_0^{1/T} d\tau\;\langle \hat{S}_{z}(\mathbf{r},\tau)\,\,\hat{S}_{z}(\mathbf{0
},0)\rangle,
\end{align}
where $\hat{S}_{z}(\mathbf{r})=\frac{1}{2}\left(\hat{n}_{\mathbf{r}\uparrow} -\hat{n}_{\mathbf{r}\downarrow}\right)$ is the $z$-component of the spin operator.
The Fourier transform of $\chispin(\mathbf{r})$, $\chispinq{q}$, diverges at the onset of long-range magnetic order of wave vector $\mathbf{q}$. As antiferromagnetic correlations are relevant near half-filling, we also consider the staggered spin susceptibility, defined in real space for $\mathbf{r}=(x,y)$ as:
\begin{align}
    \chispinstag(\mathbf{r})=(-1)^{x+y}
    \; \chispin(\mathbf{r}).
\end{align}
%
The charge susceptibility is defined by:
\begin{align}
    \chicharge(\mathbf{r})=\int_0^{1/T} d\tau\; \langle
    \delta\hat{n}(\mathbf{r},\tau)\,
    \delta\hat{n}(\mathbf{0},0)
    \rangle,
\end{align}
where $\delta\hat{n}(\mathbf{r})=\sum_\sigma \hat{n}_{\mathbf{r}\sigma}-n$, and $n$ is the average number of particles per site $n=\sum_{\sigma}\langle\hat{n}_{\mathbf{r}\sigma}\rangle$. The doping is defined as $1-n$, the deviation  from half-filling. The Fourier transform of $\chicharge(\mathbf{r})$, $\chichargeq{q}$, diverges at the onset of long-range charge order of wave vector $\mathbf{q}$.

To probe local correlations in our model we use the double occupancy: \begin{equation}
D=\langle\hat{n}_{\mathbf{r}\uparrow}\,\hat{n}_{\mathbf{r}\downarrow}\rangle,
\end{equation}
which quantifies the formation of local moments. The entrance into the Mott-insulating regime can be characterized by a plateau in the density $n$ as a function of the chemical potential $\mu$.

Alternative quantities to probe spin and charge correlations are the
equal-time structure factors:
\begin{align}
\begin{split}
    \structurespin(\mathbf{r})&=\langle\hat{S}_z(\mathbf{r})\,\hat{S}_z(\mathbf{0})\rangle\\
        \structurecharge(\mathbf{r})&=\langle \delta\hat{n}(\mathbf{r})\,\delta\hat{n}(\mathbf{0})\rangle,
     \end{split}
\end{align}
as well as their reciprocal space counterparts $\structurespin(\mathbf{q})$ and $\structurecharge(\mathbf{q})$. These have been used for this goal in previous studies~\cite{zheng2017stripe, huang2018stripe, wietekstripes2021, maier_2021_fluctuating}, along with the dynamic structure factor. However, the structure factors, while being experimentally relevant, have the disadvantage of not being directly related to ordering tendencies at finite temperature, where many phases are competing. Consequently, in this work we limit ourselves to the study of susceptibilities.

The correlation length is extracted by means of fitting momentum-space cuts of the susceptibilities using double Lorentzians with constant offsets. This procedure was also used in Ref.~\cite{huang2018stripe, maier_2021_fluctuating} for the structure factors. Unlike these previous works, we use a simple linear scale with cutoffs for our colormaps in both real and reciprocal space in order to give a fair account of the magnitude of correlations.

\subsection{Theoretical formalism}
\label{sec:formalism}
Traditional finite temperature perturbation theory is defined in the grand-canonical ensemble. More specifically, one can write a so-called bare expansion for any quantity $\mathcal{O}$ in the form of a power series of the interaction strength $U$ at fixed chemical potential $\mu$ as:
\begin{align}
    \mathcal{O}(\mu, U)=\sum_{k=0}^\infty U^k\; \mathcal{O}_k^{\text{bare}}(\mu).
\end{align}
To achieve better series convergence properties, it is advantageous to take into account the change in density  at the mean-field level
by introducing a Hartree shift to the chemical potential~\cite{rubtsov2005continuous, cdet}, which is diagrammatically equivalent to eliminating diagrams with tadpole insertions:
\begin{align}
\mathcal{O}\left(\mu_0 +  \frac{n_0(\mu_0)}{2}U,U\right)=\sum_{k=0}^\infty U^k\; \mathcal{O}_k^{\text{Hartree}}(\mu_0),
\end{align}
where $n_0(\mu_0)$ is the average number of particles per site at chemical potential $\mu_0$ and zero interaction strength.

When studying spin and charge correlations as functions of an increasing on-site repulsion $U$, it is particularly important to disentangle variations due to rather trivial changes in particle density from proper correlation effects. For this reason, we introduce in this work the fixed-density expansion: the chemical potential $\mu=\mu(n,U)$ is renormalized as a function of $U$ in such a way that the average number of particles per site, $n$, does not change with $U$:
\begin{align}
    \mathcal{O}\left(\mu(n,U), U\right) = \sum_{k=0}^\infty U^k\;
    \mathcal{O}_k^{\text{fixed-n}}(n).
\end{align}
We refer to this expansion as the ``fixed-density'' scheme, and it is the method of choice in this work, while we still sparingly use the Hartree-shifted scheme in the vicinity of half-filling and at large values of $U$, as well as for cross-checking our results. The perturbative series are numerically computed to $8-10$ expansion orders using a Monte Carlo procedure. We give more details about the formalism and the numerical methods in Sec.~\ref{sec:methods}.

\section{Results} \label{results}

\begin{figure}
\centering
\includegraphics[width=0.49\textwidth]{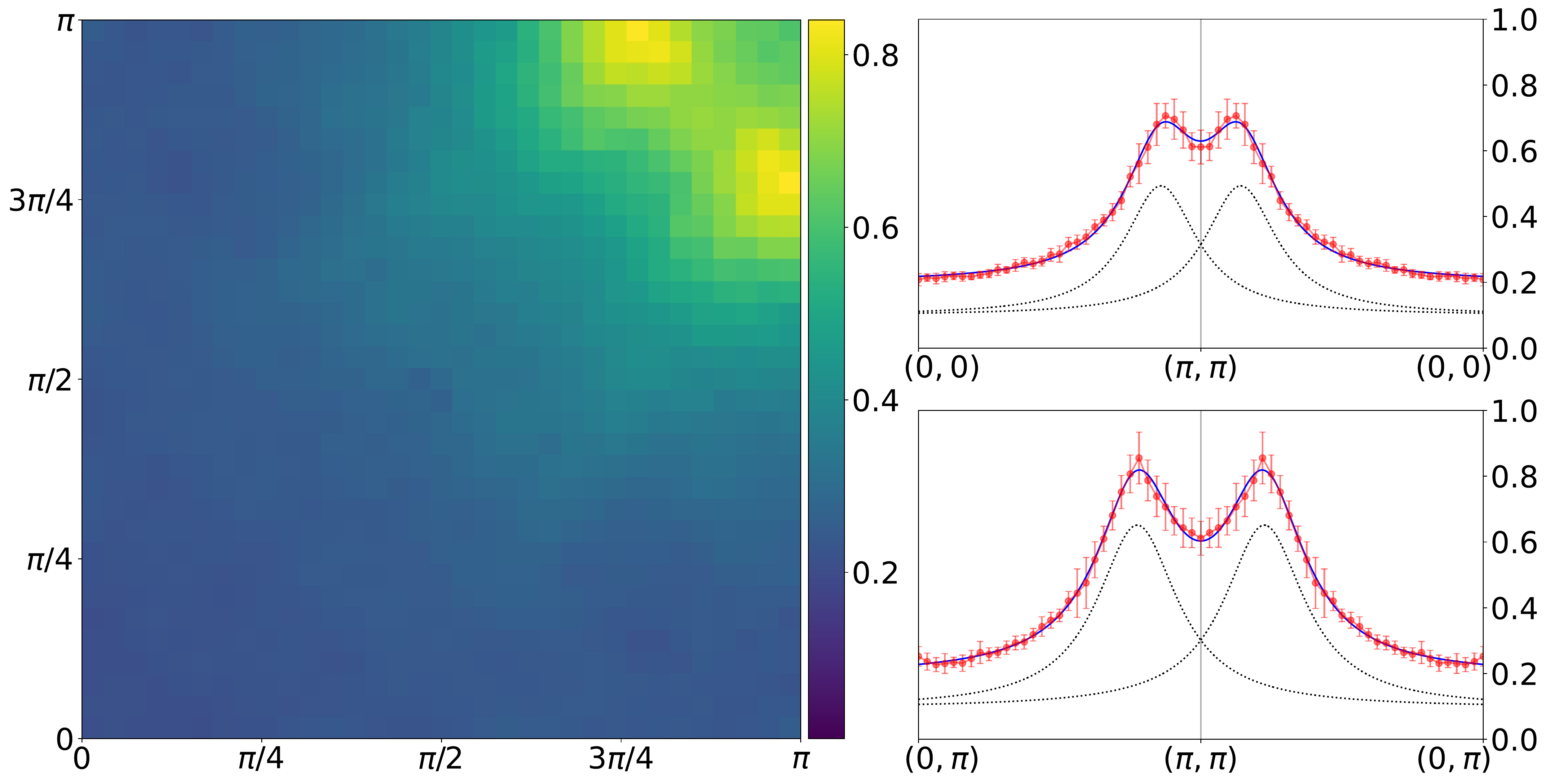}
\caption{Left: Momentum-space spin susceptibility $\chispinq{\mathbf{q}}$ for $U=5$, $T = 0.1$ and $n = 0.8$ over a quarter of the Brillouin zone. Right: Double-Lorentzian fit along the $\mathbf{q}=(Q,Q)$ (top) and $\mathbf{q}=(Q,\pi)$ (bottom) directions, where  $Q\in[0,\pi]$.
\label{Fig:chi_spin_bz}}
\end{figure}

\begin{figure}
\centering
\includegraphics[width=0.49\textwidth]{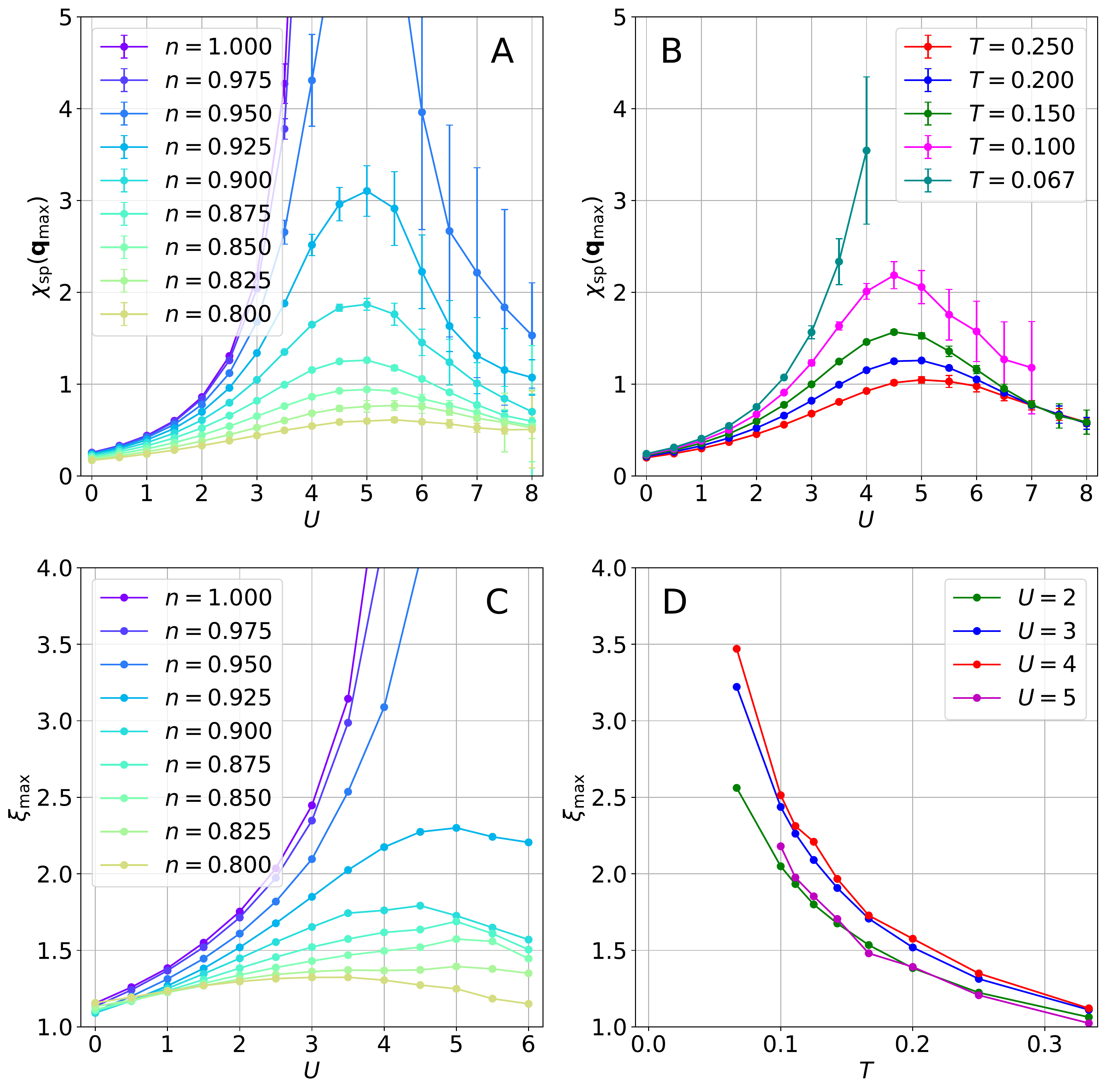}
\caption{Maximum value of the spin susceptibility across the Brillouin zone $\chispinq{q_{\text{max}}}$ (A,B) and spin correlation length $\xi$ (C,D), as a function of density $n$ (A,C), interaction $U$ (A,B,C), and temperature $T$ (D). In A and C the temperature is set to $T=0.2$. In B and D the density is set to $n=0.875$.
\label{Fig:spin_max_vs_U}
}
\end{figure}

\subsection{The three intermediate-temperature regimes of the Hubbard model}
\label{sec:stong_coupling}

\begin{figure}
\centering
\includegraphics[width=0.49\textwidth]{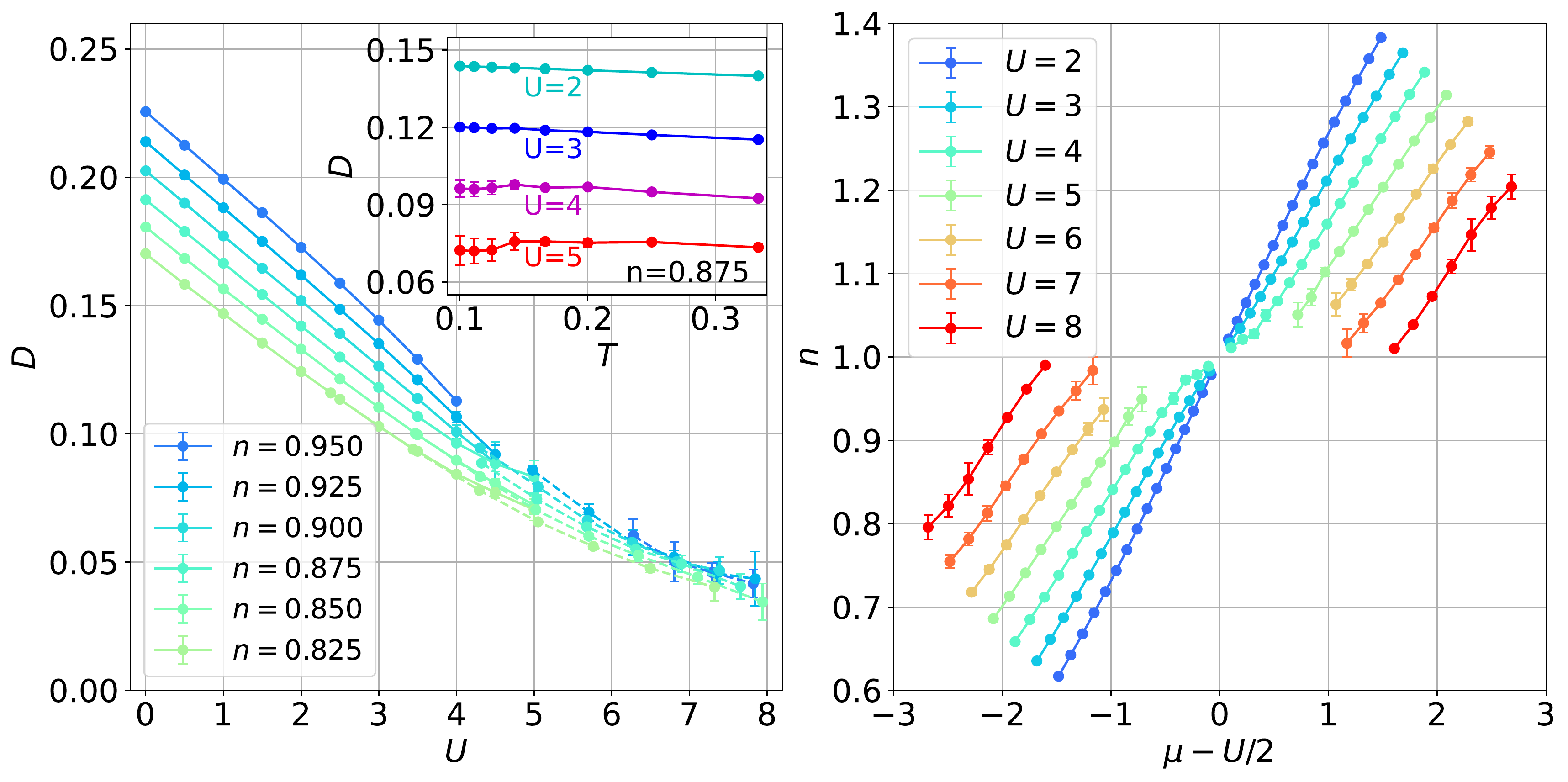}
\caption{Left: Double occupancy $D$ for $T = 0.2$ as a function of $U$ and different densities $n$. Inset: Double occupancy for density $n=0.875$ as a function of interaction strength $U$. Right: Density $n$ as a function of the chemical potential $\mu$ shifted by $U/2$ for several values of the interaction $U$.
}
\label{Fig:double_occ_mun}
\end{figure}

We commence by inspecting the
spin susceptibility $\chispinq{q}$ for all
$\mathbf{q}$ wave-vectors in the first quadrant of the Brillouin zone of a $64\times 64$ square lattice with periodic boundary conditions.
A typical result for $\chispin (\mathbf{q})$ is shown in Fig.~\ref{Fig:chi_spin_bz} for $U=5$, $T = 0.1$ and density $n=0.8$. We would like to note that no momentum interpolation procedure has been used and each point corresponds to a uniquely calculated momentum value.
For the parameters that we have investigated, the spin susceptibility
displays one or several peaks close to $\mathbf{q}=(\pi, \pi)$.
In Section~\ref{sec:commensurate_to_incommensurate} we
discuss how the location of the maxima and the associated
commensurate to incommensurate crossover depend on the
coupling $U$, the temperature $T$ and the density $n$. However, let us first focus our attention on the intensity of these peaks and the
associated correlation length $\xi$, extracted by
fitting the peaks with a double Lorentzian (plus constant offset)
Ornstein-Zernike form (see right panels of Fig.~\ref{Fig:chi_spin_bz}). \footnote{In practice we found the correlation length extracted from $(Q,\pi)$ cuts to always be slightly larger than the correlation length extracted from $(Q,Q)$ cuts for the parameters we have investigated.} The  correlation length and the maximum value of the spin susceptibility across the Brillouin zone are shown in Fig.~\ref{Fig:spin_max_vs_U} and are complemented with double occupancy and chemical potential data in
Fig.~\ref{Fig:double_occ_mun}. Together we use these quantities to identify three distinct intermediate-temperature regimes of the Hubbard model: a weak-coupling regime with short magnetic correlation length at small values of $U$, a regime with long magnetic correlation length at intermediate values of $U$ and close to half-filling, and a strong-coupling regime at large $U$ with strong short-range spin correlations (see a sketch in Fig.~\ref{Fig:phase_diagram}). We discuss these three regimes in detail below.

\subsubsection{The weak-coupling regime}

For $U \lesssim 3$ and $T\gtrsim 0.1$, the system
is in a weakly correlated regime. The double occupancy gradually decreases with increasing $U$ and slowly decreases with decreasing density, see Fig.~\ref{Fig:double_occ_mun} for
$T=0.2$. At the same time, the maximal value of the spin susceptibility as a function of $\mathbf{q}$ and the associated correlation length slowly increase with $U$ as shown in Fig.~\ref{Fig:spin_max_vs_U}.
In this regime, the double occupancy decreases with increasing temperature (see inset in Fig.~\ref{Fig:double_occ_mun})
because of the Pomeranchuk effect: an
increase of the interaction strength $U$ leads to a higher
degree of localization and a favorable gain in entropy~\cite{georges1993physical,werner2005interaction,dare2007interaction,fratino2017signatures,schaeferfate2015}.
This behaviour can be understood from the Maxwell relation $\partial S/ \partial U\big|_T = -\partial D/\partial T\big|_U$.
We observe this Pomeranchuk effect for all densities at least down to $n=0.775$.
Close to half-filling, we can observe a small
change of slope in the density versus chemical potential curve,
indicating that the compressibility at half-filling
is smaller than in the doped system, see right panel of
Fig.~\ref{Fig:double_occ_mun}.

\subsubsection{The long correlation length regime}

As the coupling $U$ is further increased, the maximal value of the spin susceptibility as a function of $\mathbf{q}$ increases rapidly, even more so as one approaches half-filling, see Fig.~\ref{Fig:spin_max_vs_U}.
The position of the maximum for $T=0.2$ is around $U \simeq 4-5$ and is only weakly dependent on
the density. This maximum is in good agreement with the prediction of Ref.~\cite{simkovicmagnetic2017}, which found the onset of magnetic spin correlations to occur at highest temperatures for $U \sim 4$. Note that this maximum is
significantly lower than the single-particle bandwidth $8t$~\cite{cheuk2016observation,parsons2016site} and seemingly slowly decreases with decreasing temperature.

In this intermediate coupling regime, the correlation length becomes
large (see Fig.~\ref{Fig:spin_max_vs_U}), especially close to half-filling and as the temperature is lowered. It then
decreases rapidly when the density is reduced and for $T \ge 0.1$ it is
only a couple of lattice sites long beyond $10\%$ doping.
As a result of this more pronounced difference in correlation
length between different density levels, the double occupancy
decreases more rapidly for occupancies closer to half-filling (see Fig.~\ref{Fig:double_occ_mun}).
This is natural as longer correlation lengths tend to favour more
localized spins.

The curves of the density versus chemical potential seem to indicate
that there is a wider region close to half-filling with a plateau
where the compressibility would be small (Fig.~\ref{Fig:double_occ_mun}).
It is however difficult to obtain results in this regime, mainly due to the
very long correlation lengths.

\begin{figure}
\centering
\includegraphics[width=0.49\textwidth]{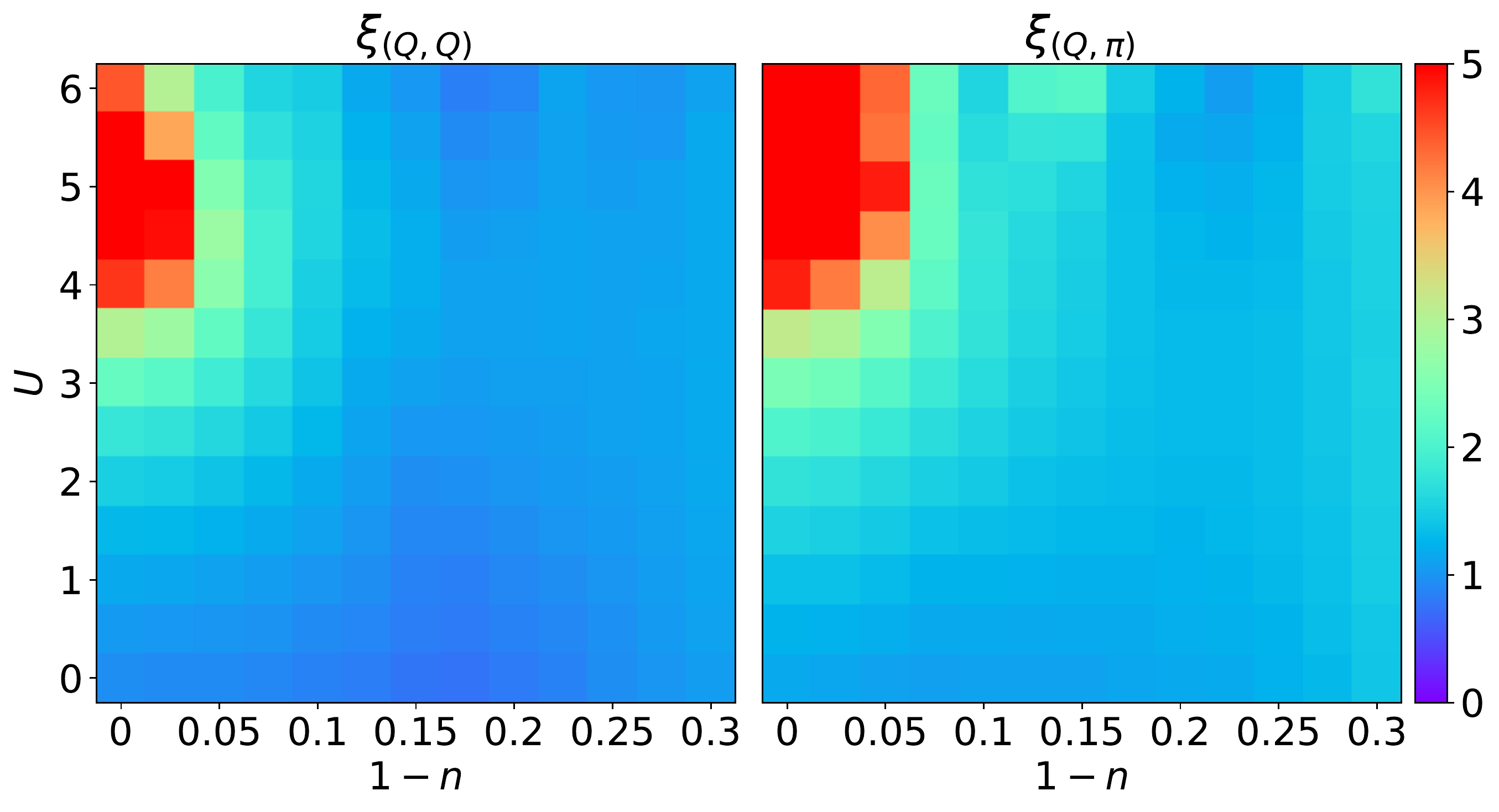}
\caption{Spin correlation length $\xi$ computed from cuts of the susceptibility along $\mathbf{q}=(Q,Q)$ (left) and $\mathbf{q}=(Q,\pi)$ (right) directions, as a function of $U$ and doping $1-n$ at fixed temperature $T = 0.2$.
\label{Fig:intensity_map_xi}}
\end{figure}

\subsubsection{The strong local correlations regime}

Finally, for $U \gtrsim 6$ we enter a regime of strong
correlations (at temperatures $T\sim0.2$). The double occupancy
that has decreased to a fraction (about $25\%$) of its non-interacting
value and has become less dependent on the density. At the same time, these (quasi) antiferromagnetic correlations are very local and
the spin susceptibility and magnetic correlation length now decrease with increasing $U$,
see Fig.~\ref{Fig:spin_max_vs_U}, and display a slower increase as
temperature is decreased as compared to the previous regime.
Also, the temperature dependence of the maximum of the spin susceptibility as a function of $\mathbf{q}$
becomes weaker.

Another clear
indication that this regime has strong correlations comes from the
behavior of the density versus chemical potential. In this regime there is a clear
plateau at half-filling showing a charge gap of the order
$\Delta \sim U/2$. This is compatible with a Mott insulating state
at half-filling~\cite{werner2007doping, wang2009anti, gull2010momentum, gull2013super, fratino2017effects}.

\subsubsection{Crossover phase diagram}

Our results for the magnetic correlation length are summarized in Fig.~\ref{Fig:intensity_map_xi} with the help of an
intensity map in the plane of doping and interaction strength as obtained from fitting $(Q,\pi)$ as well as $(Q,Q)$ momentum cuts.
Both plots clearly display a dome centered around half-filling and $U \simeq 5$ where
the correlation length is largest. The dome separates a regime of long
correlation length for $U \le 4-5$ where spin-correlation theory may provide
a reasonable description of the physics with a regime at larger $U \ge 6$
where the physics is qualitatively different and dominated by short-range
(quasi) antiferromagnetic correlations. It is interesting to note that in a recent CDet study~\cite{lenihan2020entropy} performed in the same range of parameters ($T=0.2$ and $U\lesssim 4$) a maximum in the entropy which develops away from half-filling was observed in exactly this regime of long correlations giving evidence for a crossover from metallic to non-Fermi-liquid behaviour.

From the right panel of Fig~\ref{Fig:double_occ_mun} we observe no evidence of phase separation occurring at temperature $T=0.2$ (or at $T=0.1$ which we have also investigated). The only region where we cannot make any conclusive statements is in the immediate vicinity of half-filling and at interaction strengths $U\sim 4-6$ (see also Fig.~\ref{Fig:n_mu_lines}), which has also been conjectured to be the most likely place for the occurrence of spatial phase separation by previous studies~\cite{macridin2006phase, aichhorn2007phase, chang2008spatially, galanakis2011quantum, sorella2015finite, idocompetition2018, lenihan2020entropy, sorella2021phase}.

\begin{figure}
\centering
\includegraphics[width=0.49\textwidth]{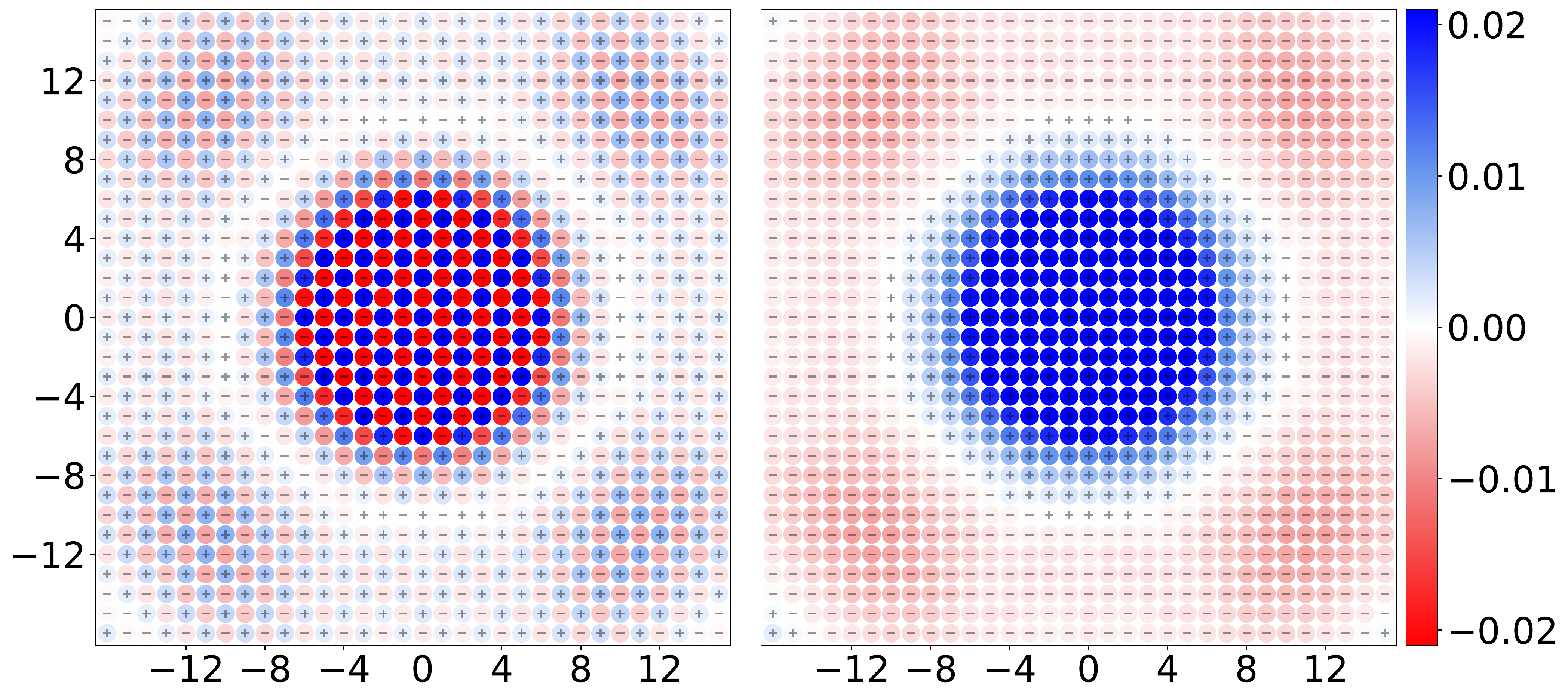}
\caption{The real-space spin susceptibility $\chispin(\mathbf{r})$ (left) and the staggered spin susceptibility $\chispinstag(\mathbf{r})  = (-1)^{x+y}
    \; \chispin(\mathbf{r})$ (right) for $U=4$, $T=0.1$, $n=0.925$.
\label{Fig:spin_r_space}
}
\end{figure}

\begin{figure*}
\centering
\includegraphics[width=0.99\textwidth]{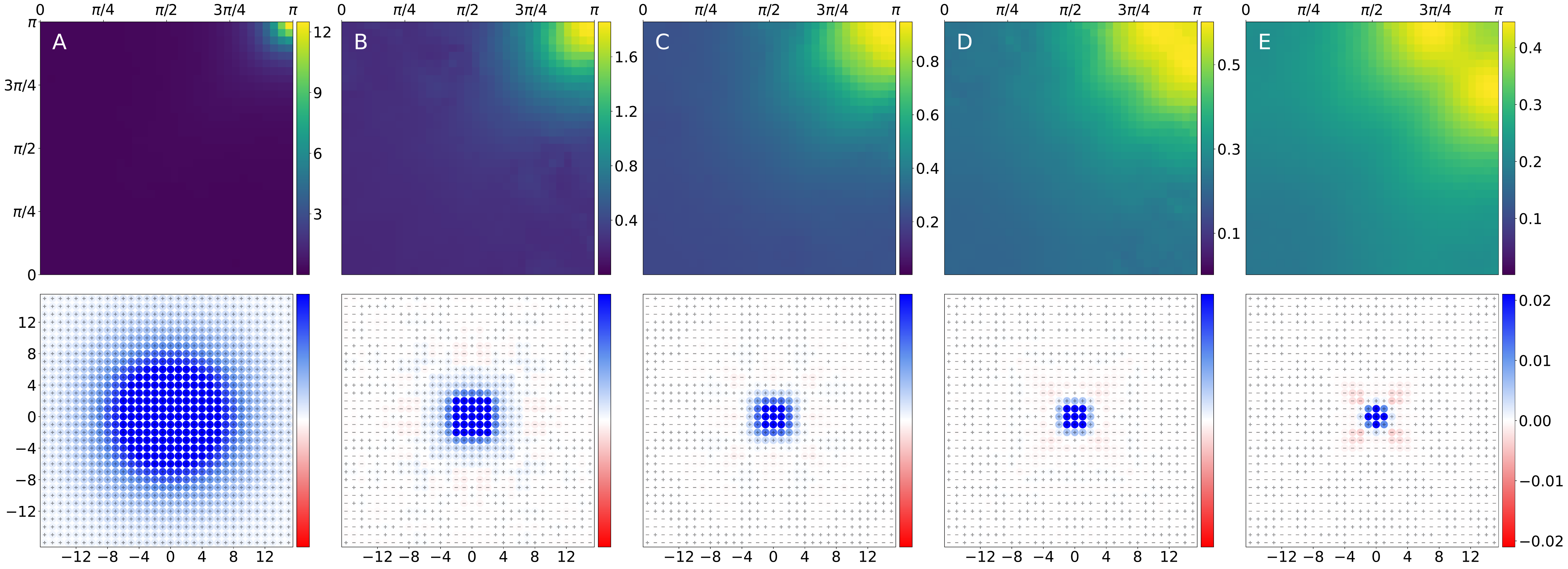}
\caption{Momentum-space spin susceptibility $\chispinq{\mathbf{q}}$ (top)
and staggered real-space spin susceptibility  $\chispinstag(\mathbf{r})$ (bottom) evaluated at $T = 0.2$ for $U=4$ at half-filling (leftmost) and for $U=5$ at densities $n = 0.9, 0.85, 0.8, 0.75$ (from left to right). We use the same colorbar scale for all real-space plots.}
\label{Fig:chi_r_q_vs_n}
\end{figure*}

\subsection{Commensurate to incommensurate spin correlation crossover}
\label{sec:commensurate_to_incommensurate}

In this section, we discuss the precise nature of the spin correlations
in regimes where the correlation length is sizeable. We systematically investigate the structure of the spin susceptibility in momentum space and the staggered spin susceptibility in real space (see Fig.~\ref{Fig:spin_r_space} for a graphical comparison to the non-staggered spin susceptibility) on large $64\times64$ lattices to study the commensurate-incommensurate crossover as a function of temperature, density and interaction strength.

Let us commence by showing the density dependence of the spin susceptibility $\chi_\mathrm{sp}$ at temperature $T = 0.2$.
We set the interaction to $U = 5$~\footnote{Note that the interaction at half-filling where $n=1$ was chosen to be $U=4$ as it wasn't possible to reliably reach $U=5$ with our method. Indeed the correlation length is expected to be prohibitively large for our algorithm in that regime.}, where we have shown
that the magnetic correlations extend furthest in real space
(see Fig.~\ref{Fig:spin_max_vs_U}). The results are shown in
Fig.~\ref{Fig:chi_r_q_vs_n}. Close to half-filling, the spin susceptibility
in momentum space is strongly peaked at $(\pi,\pi)$. Correspondingly,
in real space, the staggered susceptibility shows extended uniform antiferromagnetic correlations. As the density is reduced,
the peak first remains at $(\pi,\pi)$ and below a critical
density $n\simeq 0.95$ gradually splits into four separate peaks
at $(\pi \pm \delta_s, \pi)$ and $(\pi,\pi \pm \delta_s)$,
compatible with the square lattice symmetry.
The peaks become weaker and wider as the density is further reduced.
The resulting real-space staggered spin susceptibility displays incommensurate spin correlations for densities below $n \simeq 0.9$
with domain walls separating $\pi$-shifted regions with antiferromagnetically correlated
spins. The correlation length quickly becomes shorter
as the density is reduced.
%
%
Let us readily mention here that the onset of incommensurate
spin correlations is never accompanied by a significant redistribution
of the charge in the range of parameters that we
have studied, see Section~\ref{sec:charge}.

\begin{figure*}
\centering
\includegraphics[width=0.99\textwidth]{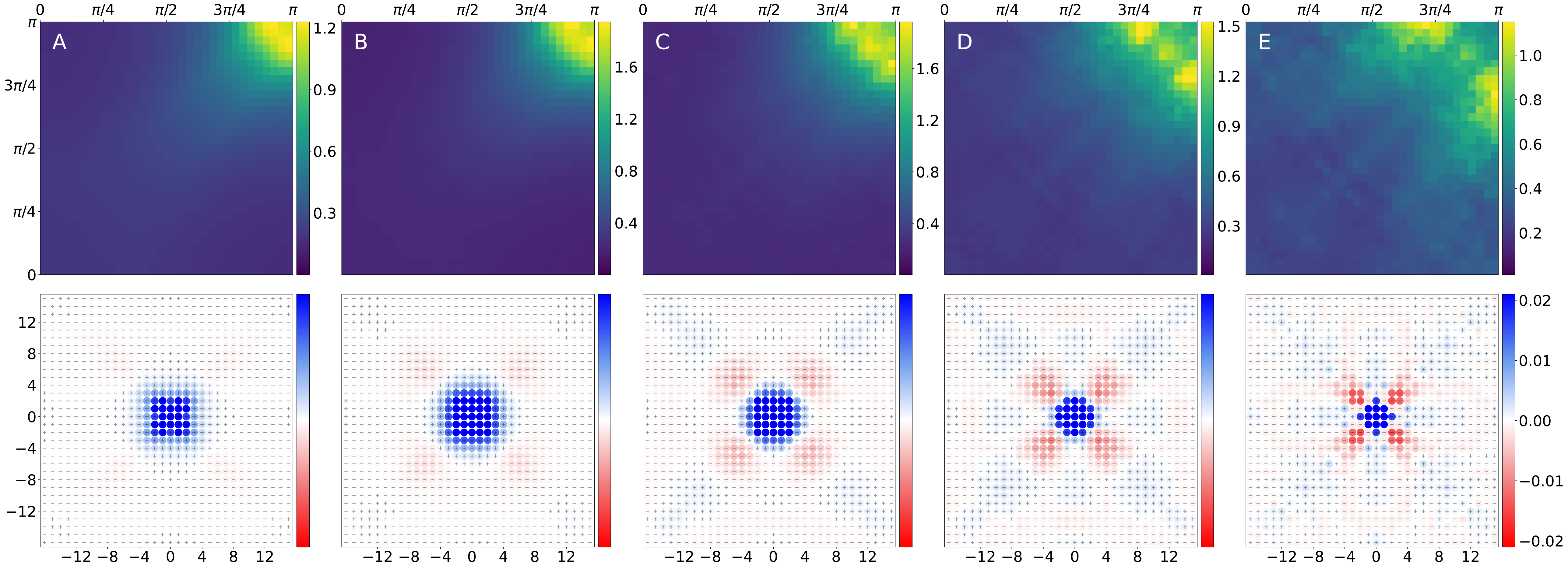}
\caption{Momentum-space spin susceptibility $\chispinq{\mathbf{q}}$ (top) and staggered real-space spin susceptibility $\chispinstag(\mathbf{r})$ (bottom) evaluated at $T=0.1$ and $n=0.875$ for $U=\{3,4,5,6,7\}$ (A-E).}
\label{Fig:chi_r_q_vs_U}
\end{figure*}

\begin{figure}
\centering
\includegraphics[width=0.49\textwidth]{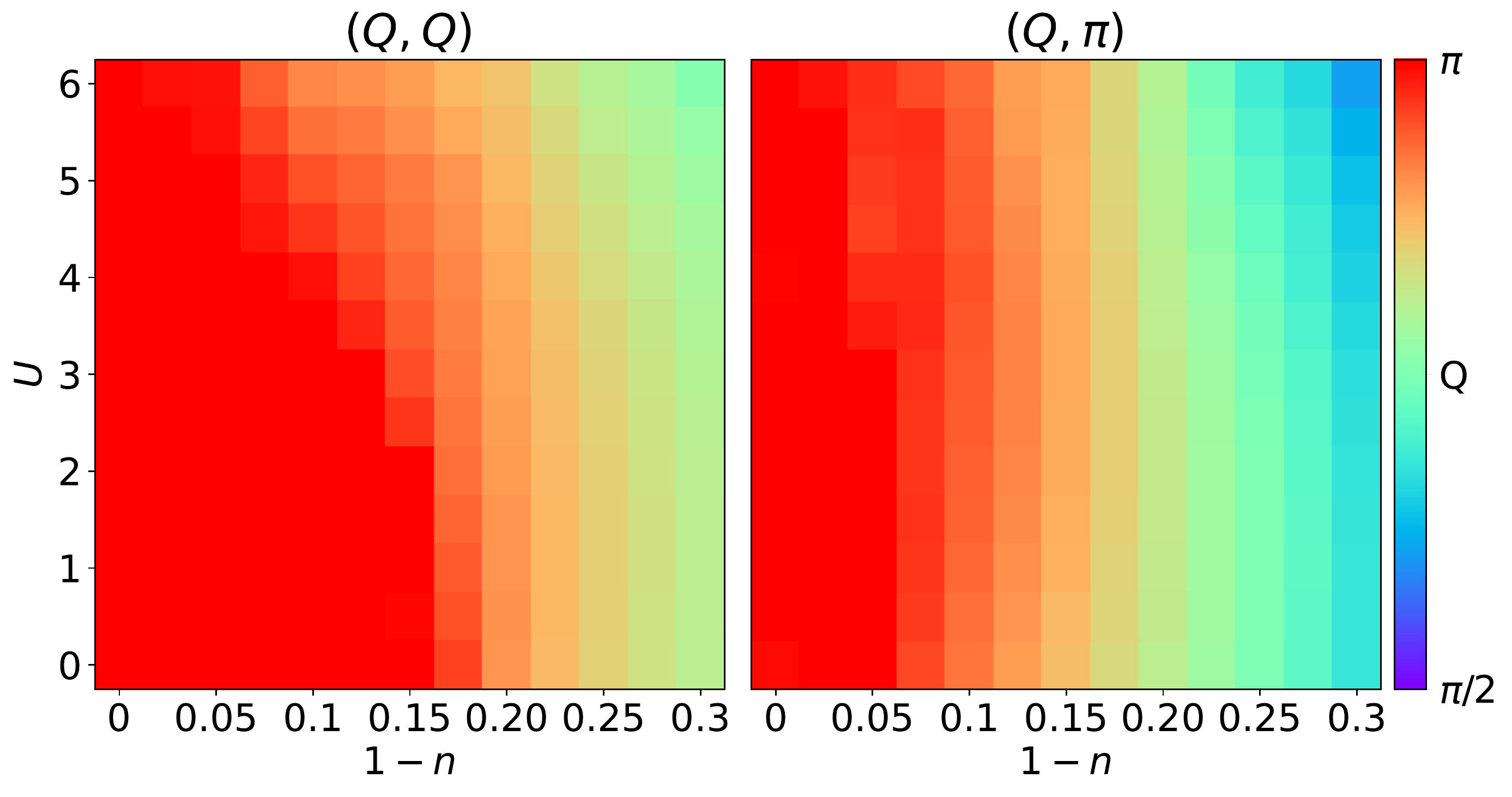}
\caption{Wave-vector component $Q$ at the maximum of the spin susceptibility along two cuts of the Brillouin zone: $Q=\argmax_{Q'}\,\chispin(\mathbf{q} = (Q',Q'))$ (left) and $Q=\argmax_{Q'}\,\chispin(\mathbf{q}=(Q',\pi))$ (right) as a function of doping $1-n$ and interaction strength $U$ for temperature $T = 0.2$.}
\label{Fig:dominant_q_vs_n_U}
\end{figure}

In Fig.~\ref{Fig:chi_r_q_vs_U} we additionally study the dependence of the crossover on interaction strength $3\leq U \leq7$ for $T=0.1$ and $n=0.875$. We find that the four peaks present at weak-coupling gradually split further away from $(\pi,\pi)$ as the interaction is increased. This corresponds to a gradually shrinking length of domain walls in real space. The increased splitting with increased interaction suggests that the leading wave-vector in the infinite interaction limit is likely different from $(\pi,\pi)$ for densities away from half-filling, as was also found in Ref.~\cite{riegler2020slave}. We also observe four sub-leading momentum-space peaks forming at vectors $(Q,Q)$ and hinting at possible diagonally-striped phases in the ground state which are commonly found in mean-field based approaches~\cite{riegler2020slave}.
We have equally found a commensurate to incommensurate crossover as a function of temperature $T$, which is described in more detail in Appendix~\ref{app:spin_susceptibility} (in particular see Fig.~\ref{Fig:chi_sp_r_q_vs_T}).

The commensurate to incommensurate crossover has previously been addressed by other numerical methods: DQMC has been used at finite temperature and predominantly on the
$16\times4$ cylinder geometry, first for the three-orbital Hubbard model~\cite{huang2017numerical} and then for the single-band Hubbard model (that we focus on in this publication) in Ref.~\cite{huang2018stripe}. The authors studied temperatures $T\geq 0.22$ for interactions in the range of $5\leq U \leq 7$. They found commensurate spin correlations in the vicinity of half-filling ($n=1$) and in the incommensurate correlations in the doped regime without next-nearest-neighbor hopping $t^\prime=0$ \footnote{The authors also studied various regimes with $t^\prime\neq0$ and found that in the hole-doped regime a negative $t^\prime$ yields incommensurate correlations while a positive $t^\prime$ yields commensurate ones.}. These findings have been confirmed by a recent DCA study on $8\times8$-sized embedded clusters in Ref.~\cite{maier_2021_fluctuating}, which was performed for $U=6$ and down to $T=0.167$. Both of these studies have only found weak hints of correlations in the charge susceptibility. Another recent study using METTS on $16\times4$ cylinders did also find incommensurate correlations at $U=10$ and $1/16$th doping ($n=0.938$), but at much lower temperatures $T<0.05$. The authors also reported sizeable corresponding charge correlations picking up below these temperatures. It is also worth mentioning that these magnetic crossovers become a actual phase transition in the three-dimensional Hubbard model, where they have been studied by means of DMFT and its diagrammatic extensions \cite{schafer2017interplay}.

Our results are summarized in Fig.~\ref{Fig:dominant_q_vs_n_U}, which shows the dominant wave-vector for different densities and values of $U$ at temperature $T=0.2$ and along the $\mathbf{q}=(Q,\pi)$ and $\mathbf{q}=(Q,Q)$ cuts. We find that the leading wave-vector is commensurate, $\mathbf{q}=(\pi,\pi)$, close to half-filling and becomes incommensurate either when the doping level is increased or when the interaction strength $U$ is larger. We also observe that the incommensurability appears earlier along the $(Q,\pi)$ than along the $(Q,Q)$ cut. In real space, we generally find that, while the leading vector is always of $(Q,\pi)$ nature, domain walls tend to form diagonally. This observation has also been made in previous studies Ref.~\cite{simkovicmagnetic2017, huang2018stripe, maier_2021_fluctuating} and was explained as the result of a superposition of a horizontal and a vertical stripe-pattern.

\subsection{Incommensurate spin correlations with no charge
redistribution}\label{sec:charge}

In this section, we want to clarify whether there is a particular
charge response connected to the onset of
incommensurate spin correlations discussed above.

Previous works~\cite{maier_2021_fluctuating, huang2018stripe, wietekstripes2021} suggest that
the formation of incommensurate spin correlations
with a wave-vector $(\pi \pm \delta_s, \pi)$
is accompanied by incommensurate charge correlations
with wave-vector $(\pm \delta_c, 0)$ where
$\delta_c \simeq 2 \delta_s$. In Ref.~\cite{huang2018stripe} no conclusive evidence of charge correlations was found at finite temperature beyond what the authors identified as boundary effects.  In~\cite{wietekstripes2021} such a $(\pm \delta_c, 0)$ peak became apparent below $T=0.05$ for $n=15/16$ and $U=10$ and for a cylindrical width-four geometry. In Ref.~\cite{maier_2021_fluctuating} the authors found a broadened maximum around the $(0,0)$ wave-vector which they fitted with a double-Lorentzian that revealed two distinct maxima at $(\pm \delta_c, 0)$. It is apparent that more conclusive evidence for such incommensurate peaks in the charge response is needed.

\begin{figure}
\centering
\includegraphics[width=0.49\textwidth]{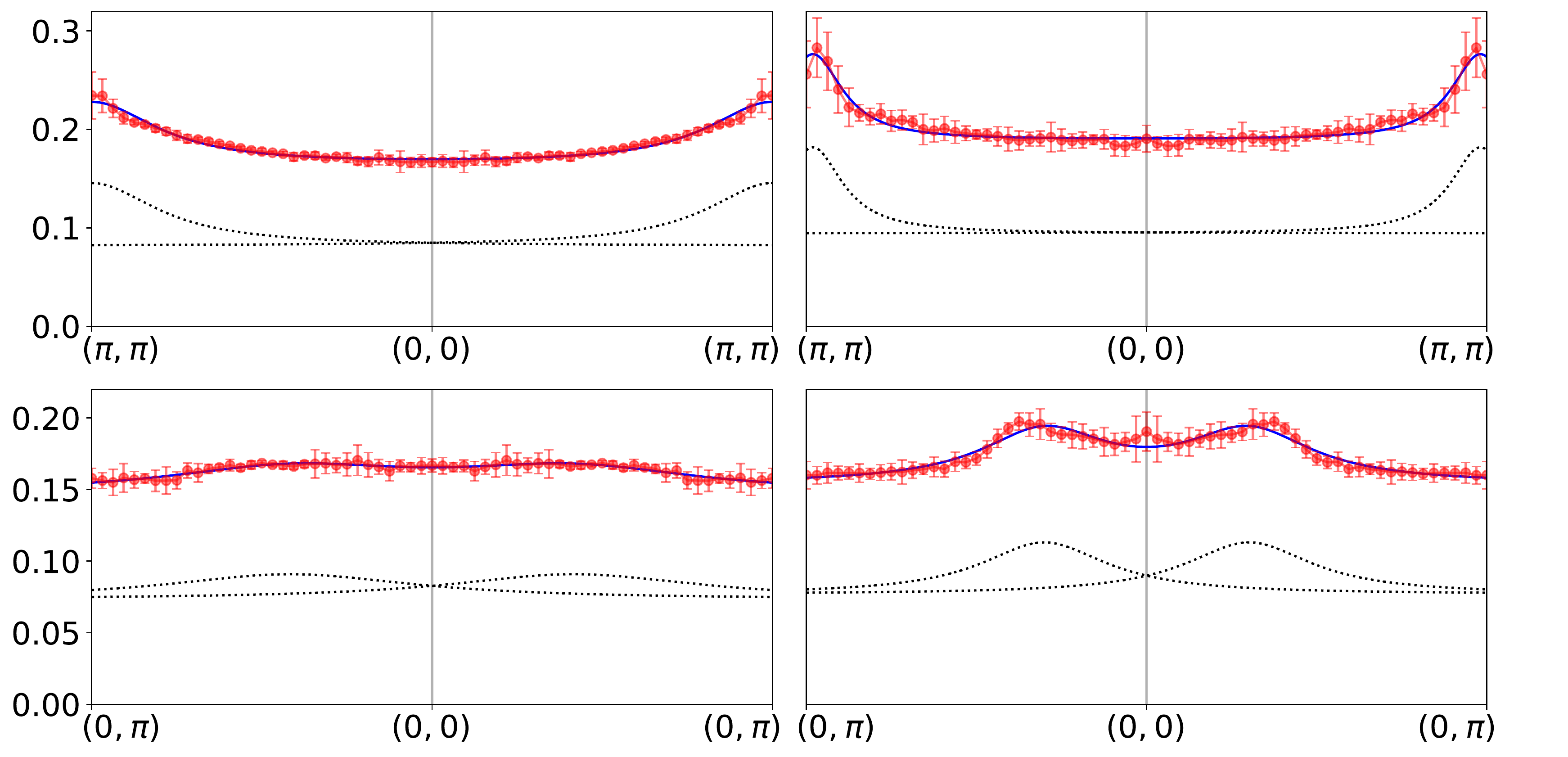}
\caption{Charge susceptibility $\chichargeq{\mathbf{q}}$ in reciprocal space. Double-Lorentzian fits along the $(Q,Q)$ (top) and $(0,Q)$ (bottom) directions. Left: $n=0.8$, $U=5$, $T=0.1$. Right: $n=0.892(6)$, $U=5$, $T=0.067$.}
\label{Fig:charge_lorentzian}
\end{figure}

In Fig.~\ref{Fig:charge_lorentzian}, we fit the zero-frequency charge susceptibility $\chicharge(\mathbf{q})$ around $(0,0)$ along the cuts $(Q,Q)$ (top) and $(0,Q)$ (bottom) using double-Lorentzians with constant offsets, much like we have done for the spin counterpart in the previous sections. In the left panel, we use the same density $n=0.8$ as Ref.~\cite{maier_2021_fluctuating} but at lower temperature $T=0.1$ and slightly lower $U=5$ (chosen to be close to the maximum of the spin correlations). Similarly to Ref.~\cite{maier_2021_fluctuating}, we find a very flat peak around $(0,0)$ which, when fitted with our procedure, reveals weak maxima at incommensurate wave-vectors $(\pm \delta_c, 0)$. In the right panel, we regard a higher density $n= 0.892(6)$ at the same $U=5$ and even lower temperature $T=0.067$, where we can more clearly identify a broad peak at $(\pm \delta_c, 0)$. The overall shape of the curve is reminiscent of what was observed in Ref.~\cite{wietekstripes2021}. We proceed to investigate how the charge response depends on the parameters of our model.

\begin{figure*}
\centering
\includegraphics[width=0.99\textwidth]{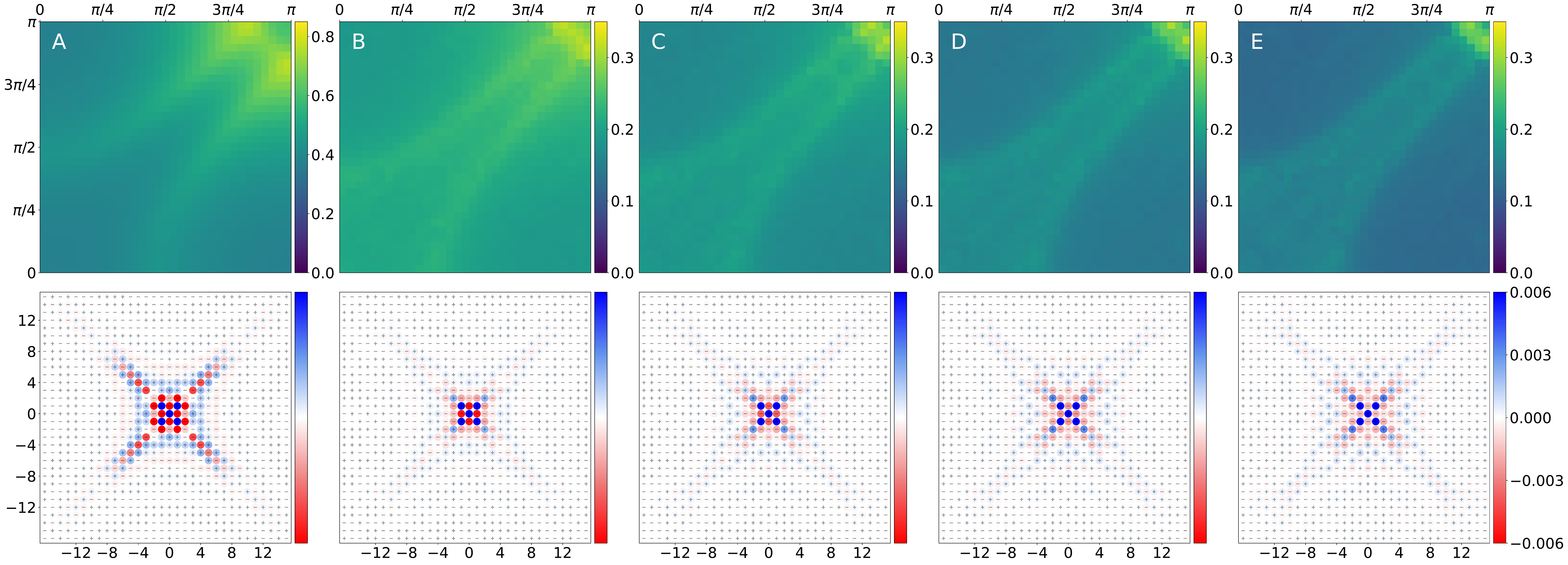}
\caption{Charge susceptibility
in reciprocal space $\chicharge(\mathbf{q})$ (top) and
in real space $\chicharge(\mathbf{r})$ (bottom) as computed from a Hartree-shifted (non-fixed density) expansion with $n_0 = 0.8$ at temperature $T = 0.067$ for interaction strengths
$U=(0,4,5,6,7)$ and densities $n=(0.8, 0.860(2), 0.892(6), 0.932(11), 0.978(20))$ (A-E).}
\label{Fig:charge_susceptibility}
\end{figure*}

\begin{figure}
\centering
\includegraphics[width=0.49\textwidth]{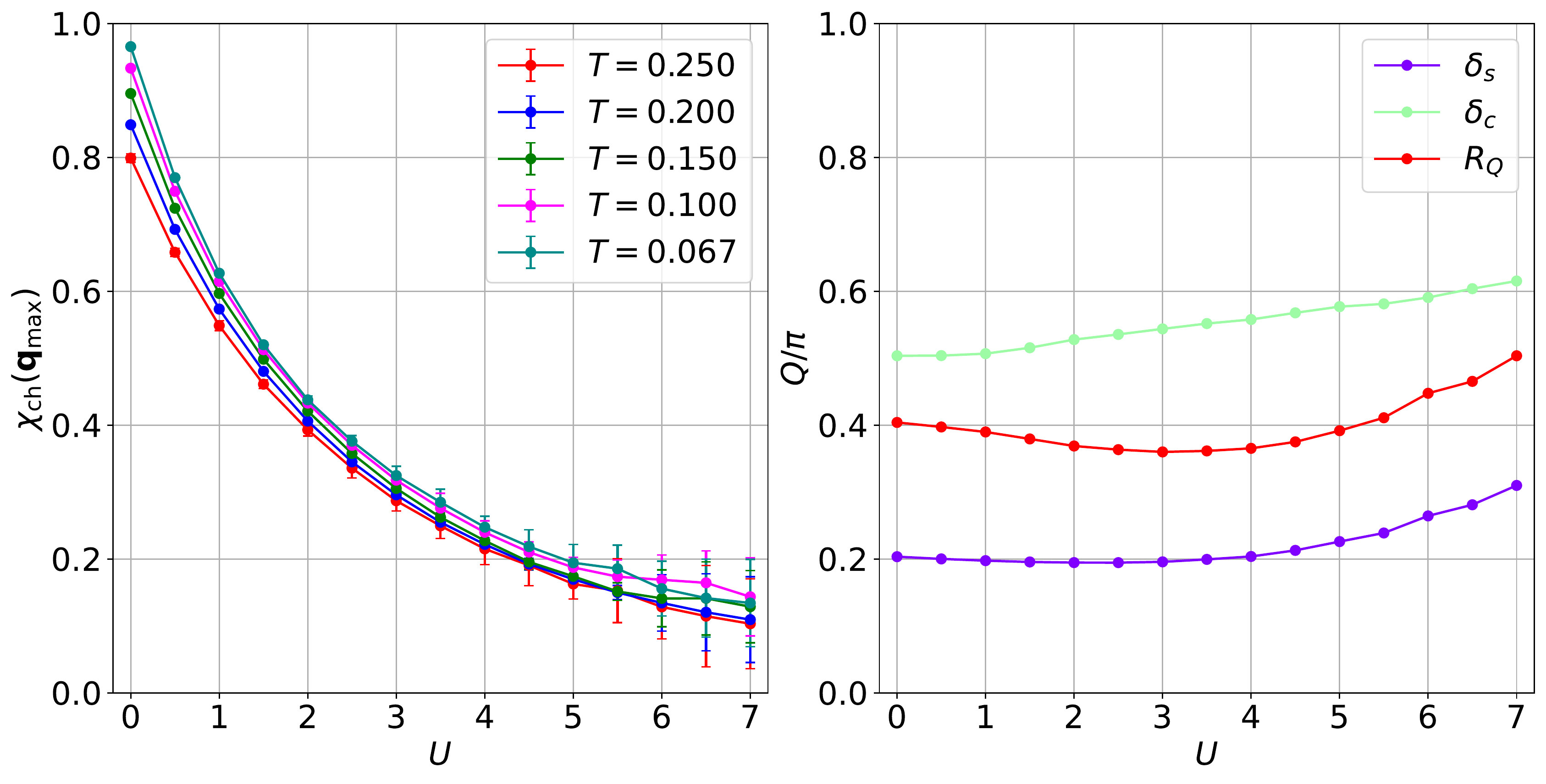}
\caption{Left: Maximum of the charge susceptibility
$\chichargeq{\mathbf{q}_{\text{max}}}$ for $n = 0.875$
as a function of the interaction strength $U$ and temperature $T$.
Right: Leading wave-vector computed from fitting with a double-Lorentzian (with constant offset) from the spin ($\delta_s$) and charge ($\delta_c$) susceptibility as well as the ratio between the two ($R_Q =\delta_s/\delta_c$). Data is shown as a function of $U$ for $n=0.8$ and $T=0.1$.}
\label{Fig:charge_susceptibility_vs_U_T}
\end{figure}

Fig.~\ref{Fig:charge_susceptibility} shows the charge susceptibility $\chicharge$
in real and reciprocal space obtained from a Hartree-shifted (non-fixed density) series expanded around the non-interacting density $n_0=0.8$ and at temperature $T=0.067$. The series is evaluated at interactions $U=\{0,4,5,6,7\}$, which corresponds to actual densities $n=\{0.8,0.860(2),0.892(6),0.932(11),0.978(20)\}$. We can identify four maxima at incommensurate wave-vectors $(\pi \pm \delta,\pi)$ along with curved broad ridge features connecting them to wave-vectors $(\pm \delta_{c},0)$. In real space we observe the formation of diagonal features which exhibit grouping of positive and negative values. It is, however, impossible to identify particular domain walls within these results.

Let us emphasize that these ridges in $\chicharge$
are characteristic of the Lindhard function and are already present in the non-interacting system~\cite{holder2012incommensurate}. They become less prominent at larger values of $U$ and wash out as the temperature in increased. For temperatures down to $T \gtrsim 0.067$ and interactions $U\lesssim 5$ at densities  $0.8 \lesssim n \lesssim 0.875$ we found a $20-25\%$ increase in the maximum of the charge susceptibility in reciprocal space from our calculations as compared to low-order RPA. At the level of RPA these ridges are determined by \emph{nesting lines} in the Brillouin zone which connect Fermi momenta with collinear Fermi velocities~\cite{holder2012incommensurate, vilardi2018dynamically}.

This absence of a significant charge response is further confirmed by studying the maximum value of the charge susceptibility
$\chicharge(\mathbf{q}_\mathrm{max})$ as a function of
$U$ and $T$ as present for density $n=0.875$ in the left panel Fig.~\ref{Fig:charge_susceptibility_vs_U_T}.
The behavior of $\chicharge(\mathbf{q}_\mathrm{max})$
is in striking contrast to that of $\chispin(\mathbf{q}_\mathrm{max})$
(see Fig.~\ref{Fig:spin_max_vs_U}). It does not display any maximum as a function of $U$ and in general has a very weak temperature dependence.

We complete this analysis by studying the position of
the leading wave-vectors given by $\delta_s$ and $\delta_c$, along the  $(Q,\pi)$ and $(Q,0)$ directions, respectively, as well as their ratio $R_Q = \delta_s / \delta_c$ at $n=0.8$ and $T=0.1$ (see right panel of Fig.~\ref{Fig:charge_susceptibility_vs_U_T}). Again, the quantities related to the spin and charge response behave qualitatively differently. Indeed,
we find that $\delta_c$ gradually increases with $U$ while $\delta_s$ stays roughly constant up until $U \sim  4-5$, which is where the strong correlation
regime starts, and then swiftly shifts further away from the $(\pi,\pi)$ wave-vector.
As a result, the ratio $R_Q$ starts decreasing and then increases beyond
$U = 4-5$ until it reaches a value close to $0.5$ at $U = 7$. While this is the
value expected for a stripe-ordered phase, we see no evidence for significant
charge correlations. An interesting question is whether the ratio remains
fixed at $0.5$ at larger values of $U$ and whether stronger charge correlations
eventually appear.

To summarize, our results support that, for the range of
parameters that we have studied, the charge correlations remain
uniform over the lattice even when longer-range spin correlations develop.
From our observations we conclude that this absence of charge stripe correlations persists in
the phase diagram at least down to $T\sim 0.10$ and for values of $U < 7$.

\begin{figure*}
\centering
\includegraphics[width=0.99\textwidth]{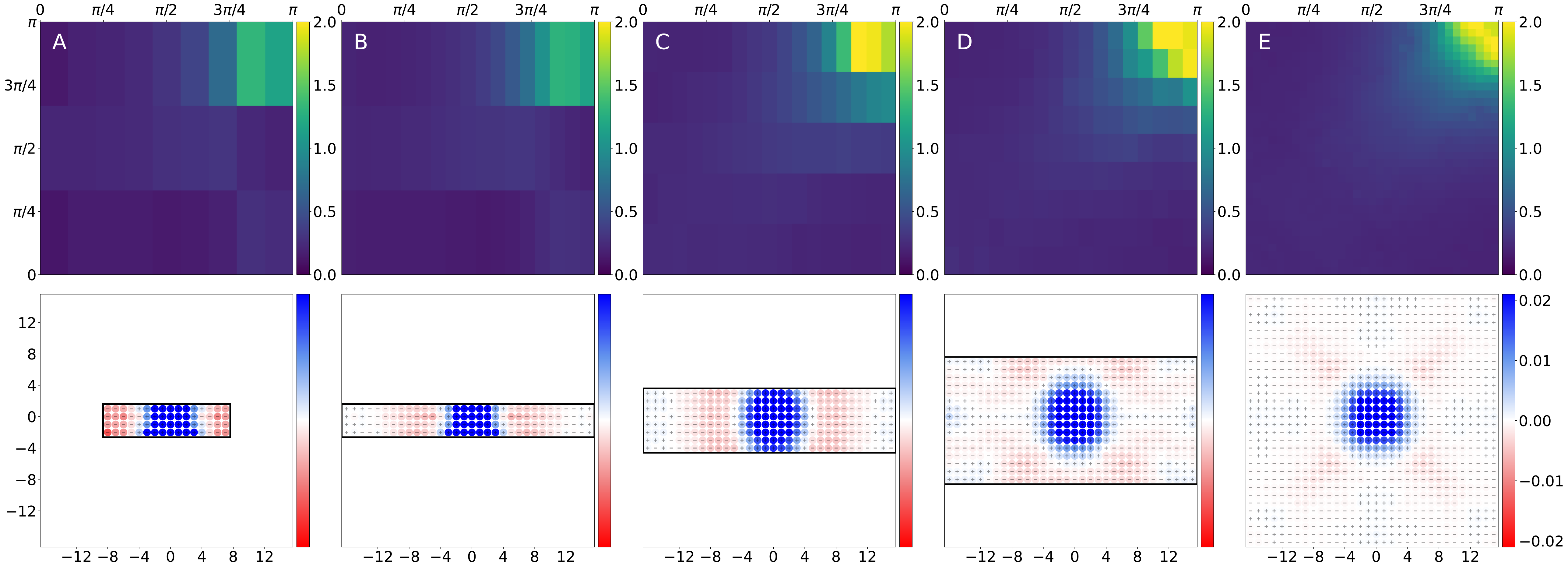}
\caption{Spin susceptibility in momentum space $\chispinq{\mathbf{q}}$ (top)
and staggered spin susceptibility in real-space $\chispinstag(\mathbf{r})$ (bottom) evaluated at $T = 0.1$ for $U=4$ and $n=0.875$ for different lattice geometries (with periodic boundary conditions) of sizes: $16\times4$~(A), $32\times4$~(B), $32\times8$~(C), $32\times16$~(D) and $64\times64$~(E). }
\label{Fig:system_size_rectangular}
\end{figure*}

\begin{figure}
\centering
\includegraphics[width=0.42\textwidth]{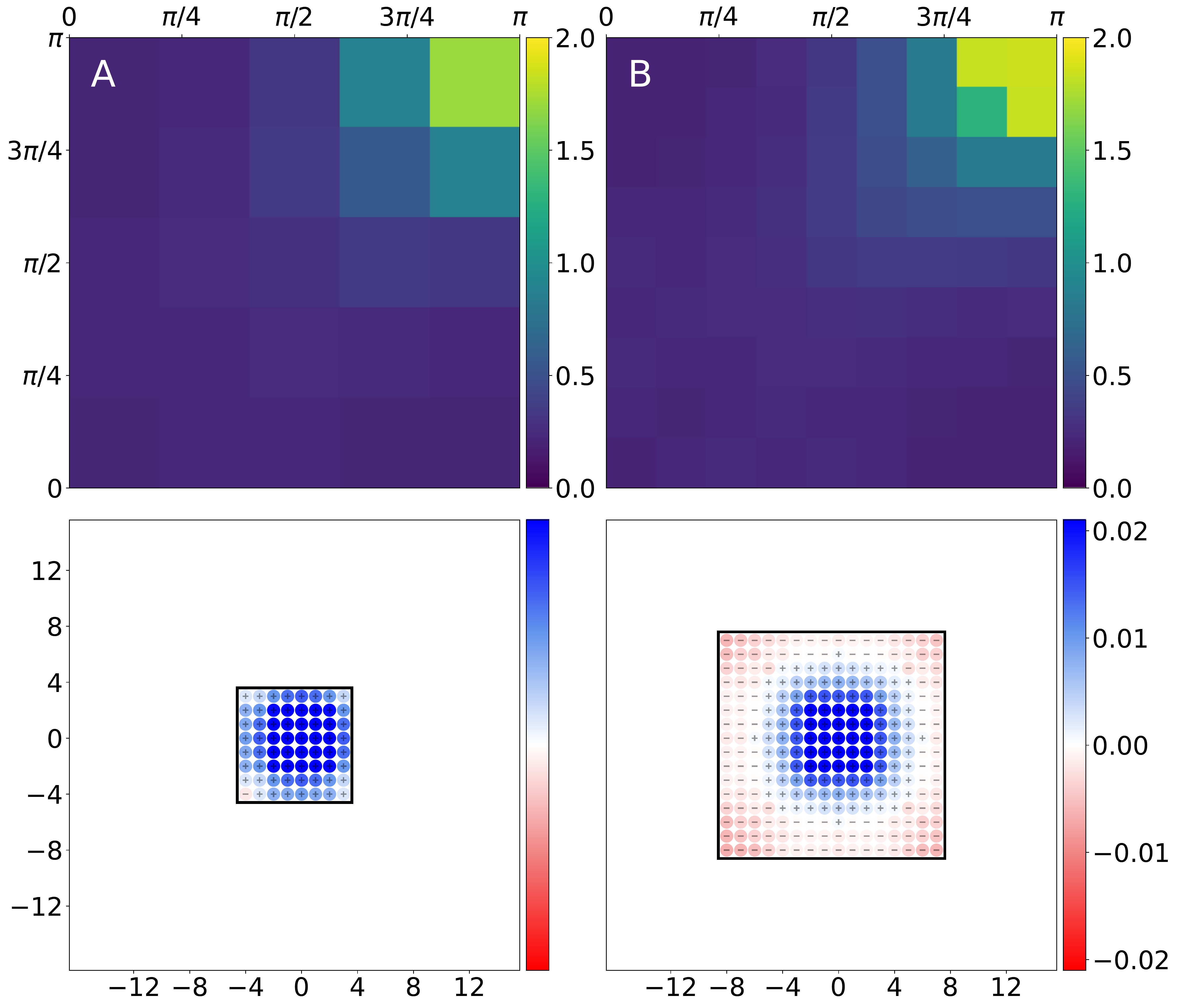}
\caption{Spin susceptibility in momentum space $\chispinq{\mathbf{q}}$ (top)
and staggered spin susceptibility in real-space $\chispinstag(\mathbf{r})$ (bottom) evaluated at $T = 0.1$ for $U=4$ and $n=0.875$ for two lattice geometries (with periodic boundary conditions) of sizes: $8\times8$~(A) and $16\times16$~(B).}
\label{Fig:system_size_square}
\end{figure}

\subsection{System size dependence}
\label{sec:system_size}
In this section, we discuss the dependence of spin and charge correlations on
the lattice size and shape. Such an analysis is important because multiple ground-state as well as finite-temperature methods can only study relatively small lattices thus making controlled extrapolations to the thermodynamic limit difficult. In particular, ground-state DMRG calculations are predominantly carried out on elongated geometries, mostly on width-four cylinders~\cite{whitestripes2003, huang2018stripe, jianground2020, qinabsence2020}, even though slightly larger width-five and width-six cylinders have also been recently investigated ~\cite{qinabsence2020}.

In contrast, a recent VAFQMC study of the ground state~\cite{sorella2021phase} has analyzed system sizes of up to $16\times16$ as well as performed an extrapolation with system size. The authors found that stripe ordered phases compete closely with superconducting ones whilst spatial phase separation also occurs for a large range of parameter space.

At finite temperatures, DQMC calculations have also been mainly performed on the $16\times4$ lattice and benchmarked against an $8\times8$ lattice geometry~\cite{huang2018stripe}. The vertical stripy patterns found for $16\times4$ has not been reproduced by the $8\times8$ lattice, which lead the authors to conclude that a square lattice should be treated as a superposition of two stripes, one horizontal and one vertical, as the rotational symmetry is not broken. Another finite-temperature method METTS~\cite{wietekstripes2021} has studied $32\times4$ sized cylindrical geometries with open boundary conditions in the long direction and periodic boundary conditions in the short one. While the method is very effective in tackling any given temperature, it is extremely hard to treat cylinders with width larger than four. It is therefore of great interest to quantify the systematics with respect to an infinite lattice. Finally, a study using DCA~\cite{maier_2021_fluctuating} has been able to inspect up to $8\times8$ clusters embedded in a bath, which the authors suggest effectively reached the thermodynamic limit for the temperatures they computed. It should be stated that despite this major technological success the momentum-resolution of such results is still limited by the size of the cluster.

In Fig.~\ref{Fig:system_size_rectangular}, we study elongated geometries, including $16\times4$ and $32\times4$, but also study $32\times8$ and $32\times16$. We use the $64\times64$ lattice as a benchmark. At density $n=0.875$ and for temperature $T=0.1 $ we chose $U=4$ which is around the maximum of correlation length. We find that both $16\times4$ and $32\times4$ qualitatively reproduce the right picture in the sense that they reveal incommensurate correlations. However, we find that the maximum in reciprocal space is underestimated by roughly $40\%$ and the length of the horizontal domain is roughly 9 sites as opposed to 13 for the thermodynamic limit result. We further see that the relative strength of highly non-local correlations gets enhanced on these cylindrical geometries. From the $32\times8$ and $32\times16$ geometries we see that even for sufficiently large geometries with broken rotational symmetry the horizontal striped patterns yield to a more complex two-dimensional pattern, which can be best described by the broadened reciprocal space peaks at $(Q,\pi)$ vectors. The fact that the real-space pattern for $32\times16$ shows the circular character found in the $64\times64$ lattice rather than well defined vertical domain walls is indicating that the system does not have a strong nematic response.

In Fig.\ref{Fig:system_size_square}, we additionally study square geometries of $8\times8$ and $16\times16$ for the same parameters. We observe that for too small lattices, such as $8\times8$ the incommensurability is effectively hidden as can be easily understood by regarding the finite momentum resolution in reciprocal space. Already for a $16\times16$ lattice the incommensurability is more apparent in real-space, while the reciprocal space plot is more inconclusive and only shows a broadened peak around the antiferromagnetic $(\pi,\pi)$ vector. All computations in this work have been done on lattices with periodic boundary conditions. We do not attempt to study the effects of the choice of different boundary conditions in this work, however such a study has been recently performed in~\cite{maier_2021_fluctuating} and considerable effects due to the choice of these have been reported.

We have also investigated the system size effects in the spin channel at higher temperatures (see Figs.~\ref{Fig:system_size_rectangular_highT},\ref{Fig:system_size_square_highT} of Appendix~\ref{app:system_size}) and system size effect in the charge channel (see Figs.~\ref{Fig:system_size_rectangular_charge},\ref{Fig:system_size_square_charge} of Appendix~\ref{app:system_size}) and found them to be a lot less severe in both cases. We conclude that finite size effects clearly worsen as magnetic correlations increase, which is the case when approaching half-filling and lowering the temperature.

\subsection{Physical insights from complex plane structure}
\label{sec:poles}

\begin{figure}
\centering
\includegraphics[width=0.45\textwidth]{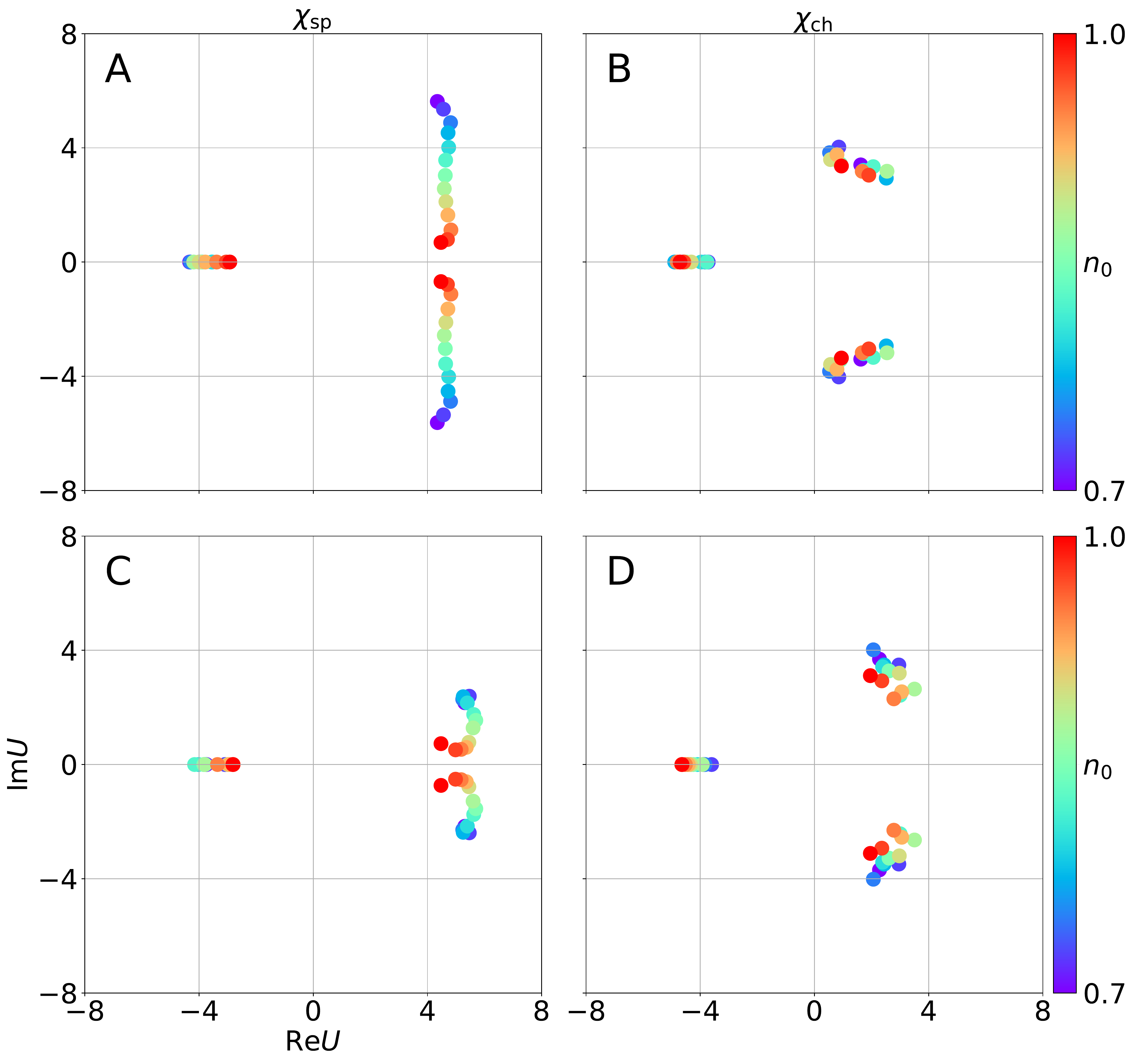}
\caption{Predicted positions of closest complex plane poles for the spin susceptibility (left) and the charge susceptibility (right) obtained at the maximum momentum wave-vector $Q$ my means of Pad\'e approximants ($[4,5]$) at temperature $T=0.2$ and as a function of the non-interacting density $n_0=n(U=0)$. Results are obtained from fixed-density series (top) and Hartree-shifted series (bottom).
}
\label{Fig:poles_fixedT}
\end{figure}

\begin{figure}
\centering
\includegraphics[width=0.45\textwidth]{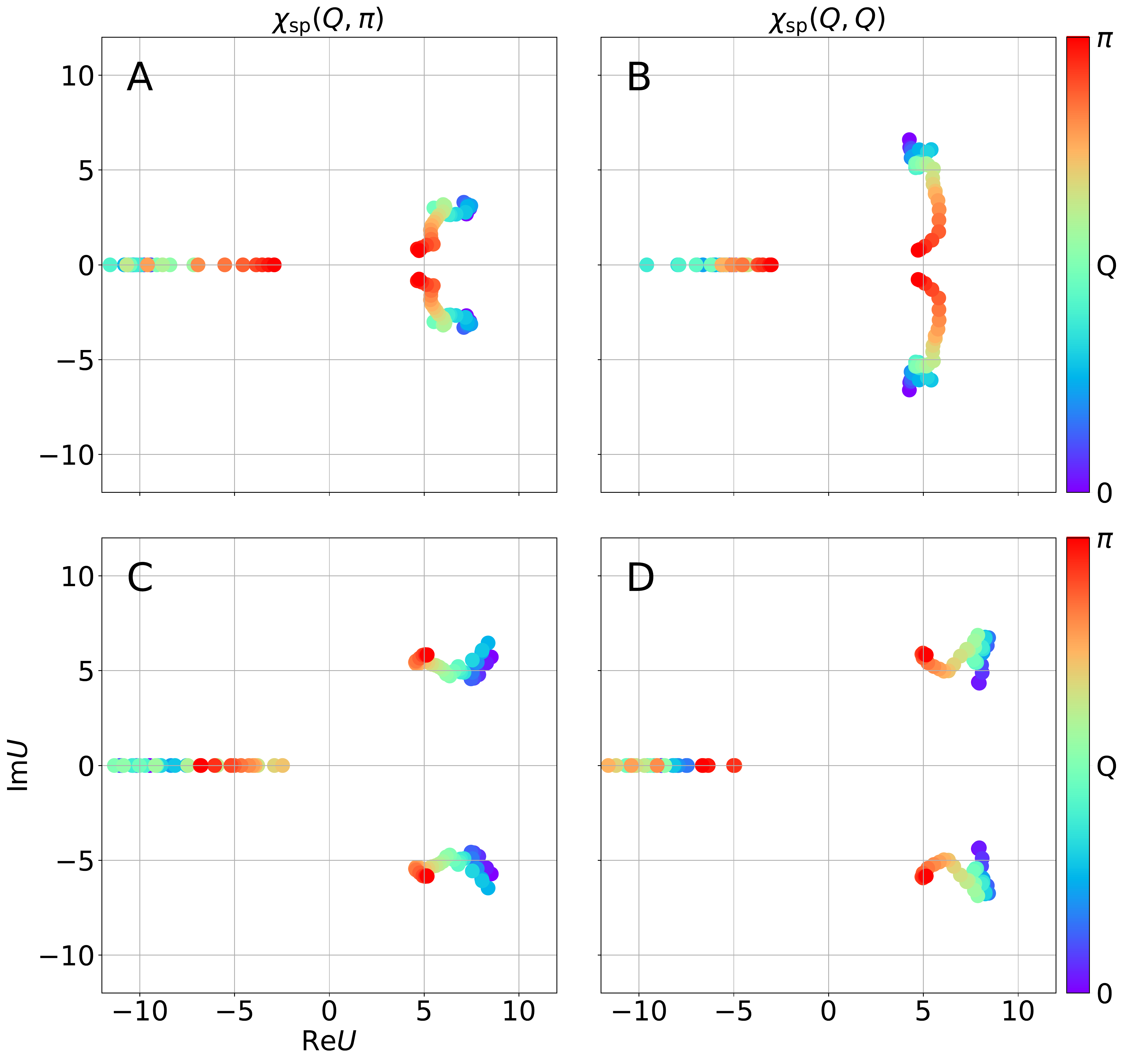}
\caption{Predicted positions of closest complex plane poles for the spin correlator obtained from Pad\'e approximants ($[4,5]$) at densities $n=0.975$ (top) and $n=0.725$ (bottom) for the momentum cuts $\mathbf{q}=(Q,\pi)$ (left) and $\mathbf{q}=(Q,Q)$ (right) as a function of $Q$, obtained from fixed-density series.}
\label{Fig:poles_fixedn}
\end{figure}

In this section, we will show that the magnetic and charge properties of
the Hubbard model can to a large extent be inferred from the
analytic structure of $\chi_\mathrm{sp}$ and $\chi_\mathrm{ch}$, seen as a function of the {\it complex} interaction strength $U$, which one can attempt to reconstruct from the knowledge of perturbative series obtained with diagrammatic Monte Carlo (see Sec.~\ref{sec:formalism}, Sec.~\ref{sec:methods} and Appendix~\ref{app:cdet} for more details on the method and formalism).

Using Padé approximants on the series coefficients, we have found the positions
of the closest poles of $\chi_\mathrm{sp}$ in the complex $U$ plane as obtained from fixed-density series within CDet.
This is shown in the top plots of Fig.~\ref{Fig:poles_fixedT}
at $T=0.2$ for $\chi_\mathrm{sp}(\mathbf{q}_{\text{max}})$. For a
given density $n$, the complex plane has a pole on the negative
real $U$ axis, most likely connected to a finite-temperature Kosterlitz-Thouless phase transition to the superconducting state of the negative-$U$ (attractive) Hubbard model \cite{paiva2004critical}. The
remaining two poles are on the positive $U$ side and
symmetrically placed about the real axis.  Quite independently from
the doping the real part of the poles is always close to
$U \simeq 5$. As a result, one expects that the spin
susceptibility will have a maximum around this value, which
is indeed what we observe, see Fig.~\ref{Fig:spin_max_vs_U}.
The poles get closer to the real axis as the density
approaches half-filling, thus yielding a larger spin susceptibility. Despite our ability to identify up to five poles in the complex plane (due to our choice of Padé approximants) the two additional poles appear significantly further away from the origin and are seemingly unstable with respect to slight changes in the series coefficients coming from a stochastic sampling of their error bars.

The right panel of Fig.~\ref{Fig:poles_fixedT}
shows the same poles for the charge susceptibility. There,
the positive $U$ poles remain further from the real axis and are much less sensitive to the choice of density $n$,
a different behavior leading to a smaller charge response
of the system. The negative pole, on the other hand is found at roughly the same values of $U$ as compared to its spin counterpart, which further corroborates the suggestion that this pole corresponds to an actual phase transition.

If we instead investigate the complex $U$ plane of the Hartree-shifted series (bottom plots of  Fig.~\ref{Fig:poles_fixedT}) we find the positive poles for the spin susceptibility to be a lot closer to the real-axis as compared to the fixed-density series poles. This is logical as the density changes with U and the series are thus sensible to regimes with very high values of the spin susceptibility in the vicinity of half-filling.

The same analysis can be performed to investigate the
the momentum dependence of the dominant peak in the spin susceptibility.
Fig.~\ref{Fig:poles_fixedn} shows the closest poles of $\chi_\mathrm{sp}(\mathbf{q}=(Q,\pi))$ and
$\chi_\mathrm{sp}(\mathbf{q}=(Q,Q))$ for two densities, $n=0.975$ (top) and $n=0.725$ (bottom). For densities in the vicinity of half-filling the closest poles are those
that have a wave-vector $\mathbf{q}=(\pi,\pi)$ in accordance with the
system developing commensurate spin correlations. As the density
is reduced, the poles move further away from the real axis but also
reorganize in a way that the closest poles to the origin are associated
with an incommensurate wave-vector $(Q,\pi)$. This makes sense since at low densities we find incommensurate correlations for practically all attainable values of $U$.

\section{Methods}
\label{sec:methods}
\subsection{Diagrammatic Monte Carlo}
In this work, we introduce and use a novel version of the diagrammatic Monte Carlo (DiagMC) method~\cite{ProkofevSvistunovPolaronShort, ProkofevSvistunovPolaronLong, vanhoucke2010, kozik2010diagrammatic, kulagin2013, vanhoucke2019, rdet, rpadet, carlstrom2021}. DiagMC has been used for lattice and continuum models with short and long-range interactions~\cite{wu_controlling, kris_felix, kun_chen, tupitsyn2017}, for interacting topological models~\cite{tupitsyn2019}, for real-time propagation~\cite{werner2009, cohen2015,gull_inchworm, olivier,  corentin,vucicevic2019real,  taheridehkordi2019algorithmic, QQMC}, and in combination with extensions of DMFT~\cite{iskakov2016, gukelberger2017,vandelli2020}. The main idea of DiagMC is to write a diagrammatic expansion for {\it intensive} physical quantities and to sample the diagrams of this expansion with a Monte Carlo procedure. As the expansion of any physical quantity can be written down directly for the thermodynamic limit, there is no overhead in considering arbitrary system sizes, thus circumventing the associated fermionic sign problem~\cite{troyer2005}. The Connected Determinant Diagrammatic Monte Carlo (CDet) algorithm~\cite{cdet} goes one step further in this direction by allowing to efficiently sum all Feynman diagrams topologies at given space-time positions of the interaction vertices, therefore reducing the computational effort.

\subsection{Efficient evaluation of chemical-potential diagrammatic insertions}
\begin{figure}
\centering
\includegraphics[width=0.47\textwidth]{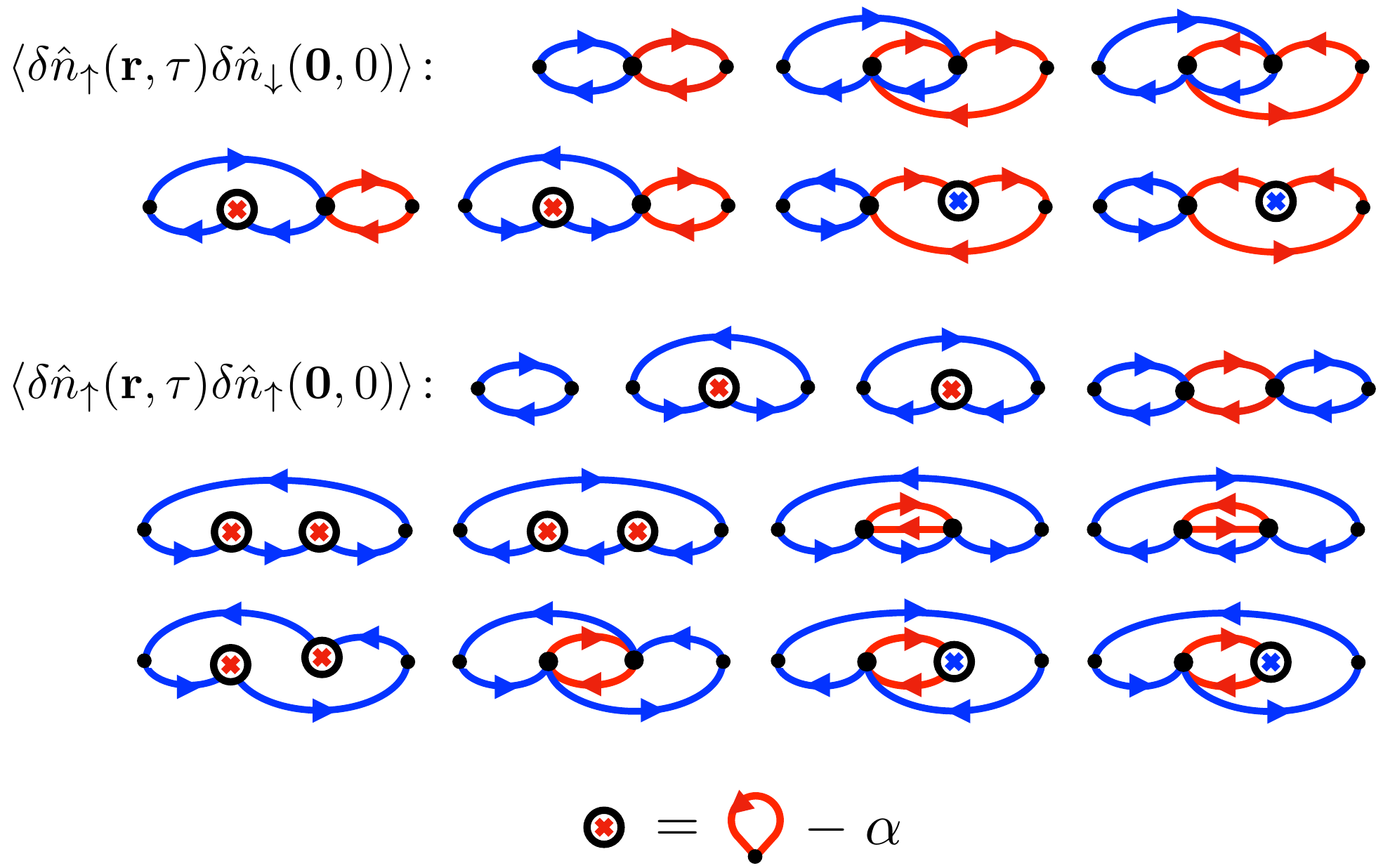}
\caption{Feynman diagrams with chemical potential insertions of the correlation functions for the spin-up and spin-down number operators, shown up to second order. Blue (red) lines represent the non-interacting $G^{(0)}$ propagators of spin up (down). The charge and spin susceptibilities are obtained from linear combinations of the integrals over imaginary-time $\tau$ of these quantities.
}
\label{Fig:feynman-diagrams-alpha}
\end{figure}

The technique we introduce in this work allows us to efficiently sum all Feynman diagrams for fixed vertex positions and arbitrary chemical-potential diagrammatic insertions, allowing in particular to fix the density perturbatively. More specifically, using the notation of Sec.~\ref{sec:formalism}, it is possible to write a double series in the interaction strength $U$ and the chemical-potential shift $\alpha\,U$
\begin{equation}
    \mathcal{O}(\mu_0+\alpha \,U,U)=\sum_{k=0}^\infty U^k \sum_{j=0}^k \alpha^j\,\mathcal{O}_{k j}.
    \label{eq:coefficients-alpha-det}
\end{equation}
 We are able to efficiently compute the coefficients of this expansion, $\mathcal{O}_{k j}$, numerically, to high order. Diagrammatically, the double $\alpha$-$U$ expansion is given by bare Feynman diagrams with chemical-potential insertions,
 as shown in Fig.~\ref{Fig:feynman-diagrams-alpha} for two building blocks of the perturbative expansions for the spin and charge susceptibilities (see Appendix~\ref{app:cdet}). The algorithm we introduce to efficiently compute these expansions is a generalization of CDet, similar in spirit to the method of Ref.~\cite{rdet}. We provide more details of our algorithm in Appendix~\ref{app:cdet}.

We can use the chemical-potential insertions to work at fixed particle number in the grand-canonical ensemble by writing the expansion for the density:
\begin{equation}
    n(\mu_0+\alpha\, U,U)=\sum_{k=0}^\infty U^k \sum_{j=0}^k \alpha^j\,n_{k j},
\end{equation}
and by solving, order by order in $U$, the following equation for $\alpha(U)$:
\begin{equation}
    n(\mu_0+\alpha(U)\,U,U) =n(\mu_0,0),
\end{equation}
where $n(\mu_0,0)\equiv n$ is a constant value and $\alpha(U)=\sum_{k=0}^\infty U^k\,\alpha_k$. The first term of the expansion for $\alpha(U)$ is given by the mean-field value $\alpha_0=n/2$, but in general there is no analytical solution for higher-order corrections, apart from half-filling where $\alpha_{k\neq 0}=0$. The computed $\alpha(U)$ is then substituted into Eq.~\eqref{eq:coefficients-alpha-det} to obtain a perturbative series in only one variable, $U$, for an arbitrary quantity $\mathcal{O}$:
\begin{equation}
    \mathcal{O}\left(\mu_0+\sum_{k=0}^{\infty}U^{k+1}\,\alpha_k,U\right)= \sum_{k=0}^\infty U^k \;\mathcal{O}_k.
    \label{Eq:single-series-final}
\end{equation}
The freedom in choosing the chemical-potential insertions $\alpha(U)$ is used in this work to cross-check the results obtained by the fixed-density expansion.

\begin{figure}
\centering
\includegraphics[width=0.49\textwidth]{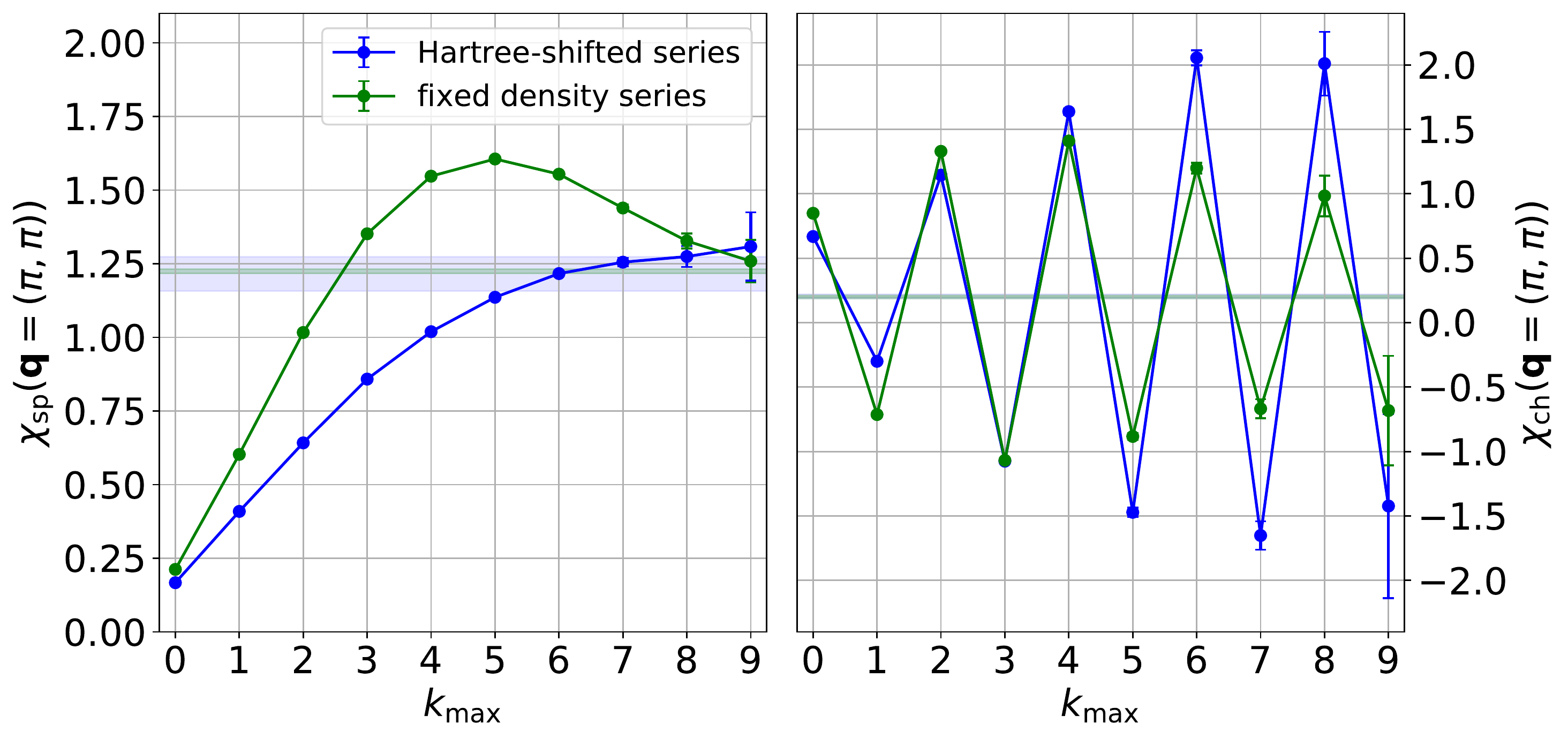}
\caption{Partial sums ($\sum_{k=0}^{k_{\text{max}}}  U^k\,\mathcal{O}_{k}$) for the spin susceptibility (left) $\chispinq{q}$ and charge susceptibility $\chichargeq{q}$ (right) at $\mathbf{q}=(\pi,\pi)$ for $n=0.875$, $U=4.3$, $T=0.2$ obtained from the fixed-density expansion (green) and the Hartree-shifted expansion
(blue).
Extrapolated results using $[4,5]$ Pad\'e approximants are shown as horizontal bands.}
\label{Fig:partial_sum_fixed_n_mu}
\end{figure}

\subsection{High-order expansions and resummation}
The series of Eq.~\eqref{Eq:single-series-final} has a non-zero radius of convergence, therefore it provides an explicit definition of $\mathcal{O}(U)$ for complex values of $U$. The radius of convergence is determined by the position of the closest singularity in the complex $U$ plane:  Fig.~\ref{Fig:poles_fixedn} and the discussion of Sec.~\ref{sec:poles} provide a physical interpretation of this otherwise purely-mathematical fact~\cite{benfatto_convergence}. For $U$ larger than the convergence radius, we can apply analytic continuation as the singularities of the susceptibilities are not on the positive $U$ axis. If that would not be the case, we would have a finite-temperature phase transition instead of a crossover, and the latter scenario is supported by the numerical data of Fig.~\ref{Fig:poles_fixedn} and physical expectations. In this work, analytic continuation is performed with the Pad\'e approximants method~\cite{gonnet2013robust}, which constitutes the only potential source of systematic errors.

The convergence and the uncertainty of our results is checked by comparing different Pad\'e schemes~\cite{fedor_sigma} to high orders (typically $8-10$), taking into account the statistical noise on our numerical estimation of $\mathcal{O}_k$, and comparing the results of the Hartree-shifted and the fixed-density expansions for the same chemical potentials. See Fig.~\ref{Fig:partial_sum_fixed_n_mu} for representative examples of partial sums for perturbative series that were obtained in our calculations.

\begin{figure}
\centering
\includegraphics[width=0.49\textwidth]{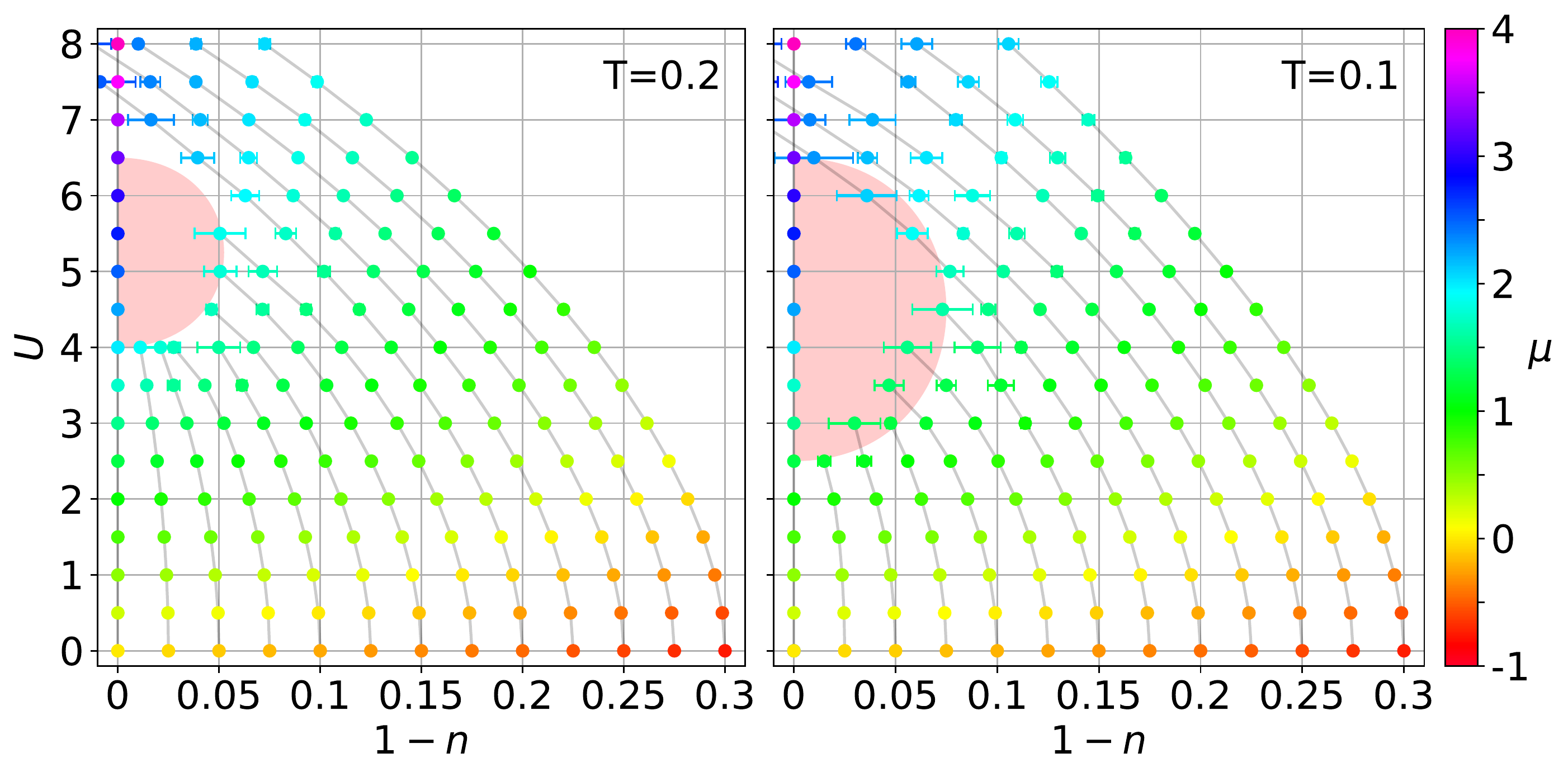}
\caption{Density $n$ as a function of $U$ computed from the Hartree-shifted series, for temperatures $T=0.2$ (left) and $T=0.1$ (right). The colorbar indicates the value of the chemical potential $\mu$. The red area shows the regime where the magnetic correlation length is large. It coincides with the parameter region where it was not possible to perform a reliable resummation of the perturbative series for the density. }
\label{Fig:n_mu_lines}
\end{figure}

We expect, on physical grounds, the region of extremely-long correlation lengths to pose a currently insurmountable obstacle to our resummation schemes, and the numerical data of Fig.~\ref{Fig:n_mu_lines} confirms this picture. This issue is common to all numerical techniques, as shown by the multi-method half-filling benchmark study~\cite{M7}. On the other hand, Fig.~\ref{Fig:n_mu_lines} also explains why previous CDet studies~\cite{cdet, fedor_sigma} were able to resum Hartree-shifted expansions far into the strong-coupling regime (up to $U=7$ for $n=0.95$ at $T=0.2$ in Ref.~\cite{fedor_sigma}) by avoiding long correlations at intermediate values of interaction strength $U$.

\subsection{Monte Carlo evaluation of the integrals}
\label{sec:diagmc}
The coefficients of Eq.~\eqref{eq:coefficients-alpha-det} are computed from the integral of the sum of all Feynman diagrams at fixed space-time vertex positions:
\begin{equation}
    \mathcal{O}_{kj}=\frac{1}{k!}\int_{\tau_1,\dots,\tau_k}\sum_{\mathbf{r}_1,\dots,\mathbf{r}_k} O_{j}(\{(\mathbf{r}_1,\tau_1),\dots,(\mathbf{r}_k,\tau_k)\}),
    \label{eq:coefficients-integral}
\end{equation}
where $\mathbf{r}_l$ is a lattice site and $\mathcal{\tau}_l\in\left[0,\frac{1}{T}\right]$ is the imaginary-time. The integrals of Eq.~\eqref{eq:coefficients-integral} are evaluated with the Many-Configuration Markov-chain Monte Carlo method introduced in Ref.~\cite{MCMCMC} and determinants are evaluated using a fast principal minor algorithm~\cite{griffin2006principal,vsimkovic2021fast}. As briefly mentioned before, if the quantity $\mathcal{O}$ is well-defined in the thermodynamic limit, its series coefficients $\mathcal{O}_{kj}$ are also well-defined, and there is no problem to extend the integration range of Eq.~\ref{eq:coefficients-integral} to larger system sizes, or even directly to the thermodynamic limit.

\section{Conclusions} \label{conclusions}

In this paper, we have established numerically-exact results for the momentum-resolved charge and spin susceptibilities of the two-dimensional Fermi-Hubbard model. We have systematically studied
a range of couplings $0 \le U \le 8$, densities $0.7 \le n \le 1.0$ and
temperatures down to $T = 0.067$. We have shown that there are three distinct
regimes separated by rather sharp crossovers:
A weak-coupling regime with short magnetic correlation length,
an intermediate coupling regime where the correlation length rapidly increases
when the temperature is lowered or when the density approaches half-filling,
and a strong coupling regime with short-range magnetic correlations.
These regimes are also characterized by different behaviors of the double
occupancy. In the weak coupling regime, the double occupancy decreases with
increasing temperature, displaying a Pomeranchuk effect. At strong coupling instead,
the electrons are more localized and a Mott gap appears at half-filling.

In the regime with long magnetic correlation length, we established the existence
of a crossover from $\mathbf{q}=(\pi, \pi)$ antiferromagnetic correlations to incommensurate correlations. This crossover is mainly seen as a function of increasing doping, but also as temperature is lowered or as the interaction is increased. The most common incommensurate vectors are $(\pi, \pi \pm \delta q)$, $(\pi \pm \delta q, \pi)$ but we found hints of the existence of peaks at $(\pi \pm \delta q, \pi \pm \delta q)$, which we expect to become stronger in parts of the phase diagram at lower temperatures.
We have shown that the different phenomena we observed can, to a large extent, be
read from the analytic structure of $\chi_\mathrm{sp}$ and $\chi_\mathrm{ch}$
as function of a {\it complex} interaction strength.

In the parameter range that we investigated, we found that the charge susceptibility
behaves very differently from the spin susceptibility and saw no evidence
of a reinforcement of the charge response in regimes where the
(incommensurate) magnetic correlation
length increases rapidly. We attribute the existence of weak peaks
at $( \pm \delta q_c, 0)$ to the presence of similar peaks at the non-interacting level
in the Lindhard function. This absence of charge stripe correlations above
$T \simeq 0.067$ raises interesting questions about the conciliation of the
finite temperature properties of the system and its ground-state properties~\cite{zheng2017stripe, qinabsence2020}.
A natural scenario would be that, as temperature is lowered further,
charge-stripe correlations eventually appear, as this is what happens in the width-four cylinder geometry~\cite{wietekstripes2021}, but further work is needed to confirm this picture.
An additional question is whether such an ordered stripe phase would appear at finite temperature or only exists in the ground state.

Our findings from benchmarking different lattice geometries show that,
especially as temperature is lowered and correlations increase, finite-size effects do lead to
sizeable quantitative, and in the case of very small lattice sizes even quantitative, differences in results. It would be interesting to understand how these effects compare to the small energy differences that are observed between different competing ordered ground-state phases that have been studied on width-four cylinders and the $16\times16$ lattice~\cite{whitestripes2003, huang2018stripe, jianground2020, huang2018stripe, sorella2021phase}.

In this work, we have limited ourselves to the discussion of the Hubbard model without next-nearest-neighbor hopping $t'$, which is however believed to play a crucial role in being able to connect the model to experimental findings on cuprate superconductors with effective $t'$ in the range of $-0.4 \leq t^\prime \leq -0.1$~\cite{pavarini2001band}. Indeed, there has been an extensive amount of neutron scattering experiments performed on cuprates studying commensurate and incommensurate magnetic correlations~\cite{yamadadoping1998, yamadacommensurate2003,armitageprogress2010, fujitastrip12004, enokispin2013, tranquada1995evidence,tranquada2004quantum, hayden2004structure, comin2016resonant}. In particular, it was found that electron doped cuprates show commensurate peaks~\cite{yamadacommensurate2003,armitageprogress2010}, while experiments on hole-doped cuprates found the peaks to be incommensurate~\cite{tranquada1995evidence,tranquada2004quantum, hayden2004structure}. Another interesting difference was found between La-based cuprates and LBCO, where for the former both charge and spin ordering vectors increase with doping, while for the latter the charge vector decreases and the spin one increases. For the $t^\prime=0$ case studied here we find behaviours consistent with La-based cuprates, similarly to Ref.~\cite{huang2018stripe}. It is, of course, impossible to talk about cuprates without mentioning superconductivity and its interplay with magnetic order. In that respect it would be of great interest to study superconducting $d$-wave correlations on equal footing with spin and charge ones, much like what has been recently carried out in Ref.~\cite{maier_2021_fluctuating}. It would be
equally important to relate our findings with single-particle properties and to explore a possible connection to pseudogap physics. We, however, leave these investigations for future work.

The results obtained in this work were, to a great degree, enabled by algorithmic progress, in particular the development of a connected determinant diagrammatic Monte Carlo algorithm for double expansions, in $U$ and an arbitrary chemical potential shift $\alpha(U)$. Importantly, this allowed us to make the choice of the function $\alpha(U)$ \emph{a posteriori} and without any additional computational effort. Here, we have limited ourselves to only two such choices, the Hartree-series and the fixed-density series. However, one can generate an arbitrary number of alternative expansions from different choices of $\alpha(U)$. This, on one hand, provides an additional degree of cross-control over the systematics coming from the resummations of perturbative series. On the other hand, this freedom of choice can potentially be exploited through optimisation schemes such as machine learning based algorithms.

\section*{Acknowledgements} \label{acknowledgements}
The authors thank Antoine Georges, Hanhim Kang and Thomas Schäfer for valuable discussions. This work was granted access to the HPC resources of TGCC and IDRIS under the allocations A0090510609 attributed by GENCI (Grand Equipement National de Calcul Intensif). This work has been supported by the Simons Foundation within the Many Electron Collaboration framework.

\appendix




\begin{figure*}
\centering
\includegraphics[width=0.99\textwidth]{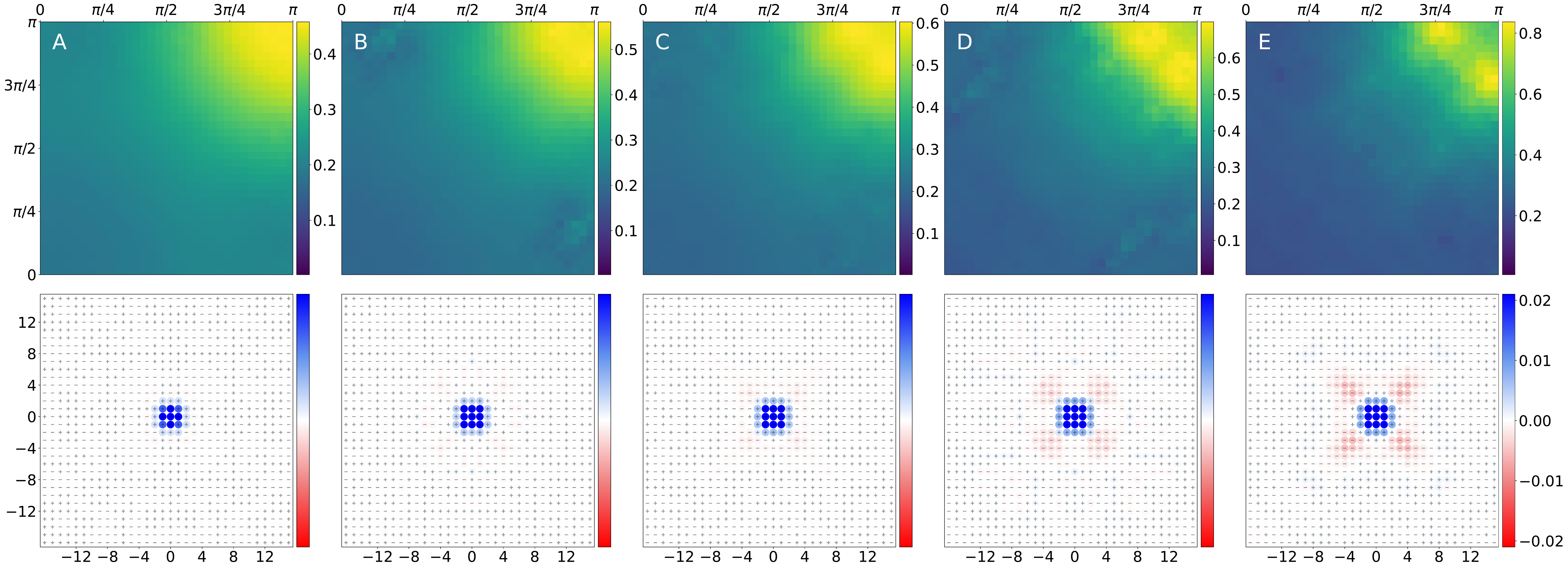}
\caption{Momentum-space spin susceptibility $\chispin(\mathbf{q})$ (top) and real-space staggered spin susceptibility $\chispinstag(\mathbf{r})$ (bottom) evaluated at $U=5$ and $n=0.8$ for temperatures $T=\{ 0.33, 0.25, 0.20, 0.167, 0.10 \}$ (A-E).}
\label{Fig:chi_sp_r_q_vs_T}
\end{figure*}

\begin{figure*}
\centering
\includegraphics[width=0.99\textwidth]{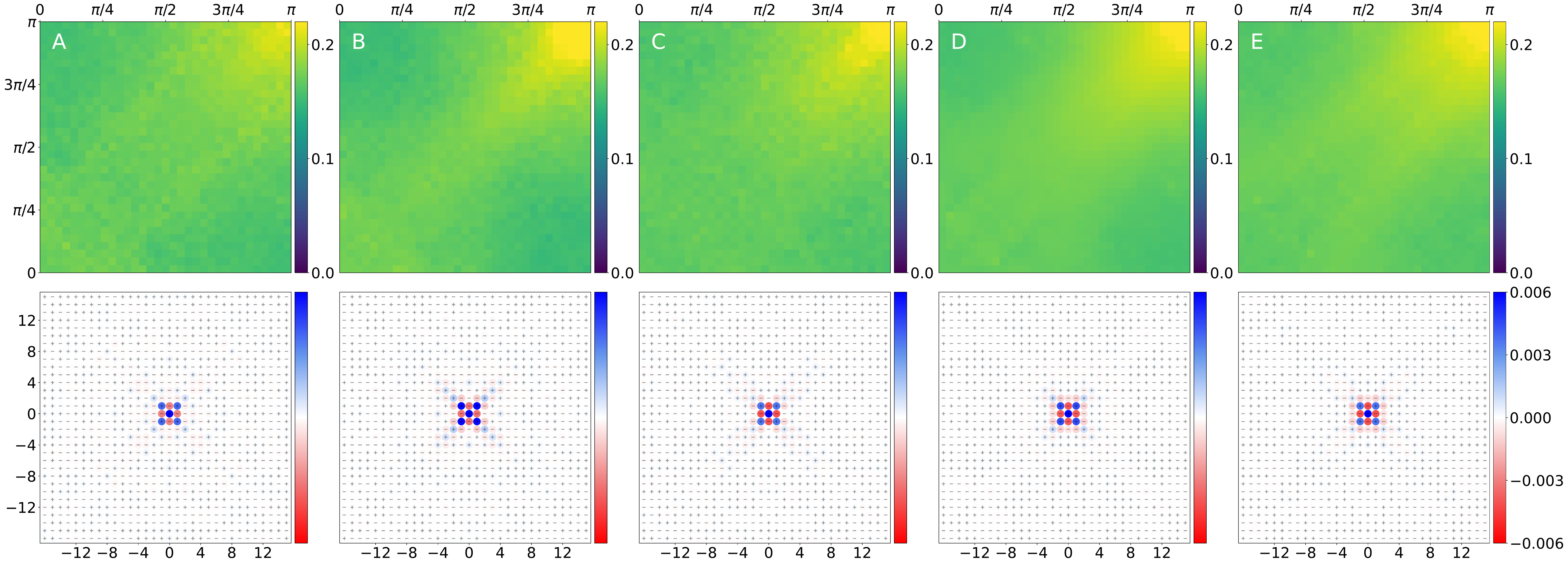}
\caption{Charge susceptibility in momentum space $\chi_\mathrm{ch}(\mathbf{q})$ (top) and  in real space $\chi_\mathrm{ch}(\mathbf{r})$ (bottom) evaluated at $U=5$ and $n=0.8$ for temperatures $T=\{ 0.20, 0.167, 0.125, 0.10, 0.067 \}$ (A-E).}
\label{Fig:chi_ch_r_q_vs_T}
\end{figure*}

\section{Spin and charge susceptibilities as a function of temperature}
\label{app:spin_susceptibility}

Here, we have conducted the commensurate to incommensurate crossover analysis at a fixed density $n = 0.8$, interaction strength $U=5$ and for different temperatures $T$, see
Fig.~\ref{Fig:chi_sp_r_q_vs_T} and Fig.~\ref{Fig:chi_ch_r_q_vs_T}.

The spin susceptibility (see Fig.~\ref{Fig:chi_sp_r_q_vs_T}) in reciprocal space $\chi_\mathrm{sp}(\mathbf{q})$ has a broad peak is centered around wave-vector $(\pi,\pi)$ at temperatures above $T \simeq 0.3$. At lower temperatures, the broad peak splits into four sharper peaks at $(\pi, \pi \pm \delta_s)$, $(\pi \pm \delta_s, \pi)$. The staggered real space spin susceptibility $\chi_\mathrm{sp}(\mathbf{r})$ meanwhile builds up corresponding diagonal domain walls across the crossover. The length of the domain around the center stays relatively unaffected by temperature at 5 lattice sites.

We also study the charge susceptibility
in real and reciprocal space for the same set of parameters, where the spin correlations clearly undergo a commensurate to incommensurate crossover (see Fig.~\ref{Fig:chi_ch_r_q_vs_T}). The reciprocal space charge susceptibility $\chi_\mathrm{ch}(\mathbf{q})$ shows little structure except for a clear peak around $(\pi,\pi)$. At the lowest available temperature $T=0.067$ we observe the start of the formation of ridges as discussed in relation with Fig.~\ref{Fig:charge_susceptibility}, as well as the corresponding weak maxima at wave-vectors $(\delta_c,0)$ and $(0,\delta_c)$. The real space charge susceptibility $\chi_\mathrm{ch}(\mathbf{r})$ always takes positive value at $\mathbf{r}=(0,0)$, which is expected because because the system is hole-doped $n<1$ and the double-occupancy must be positive $D>0$. There is no indication of domain walls forming, where the
correlator would change sign as seen for the spin
susceptibility for these parameters. We observe a similar behavior for all the temperatures that we have investigated here
$T\in\{0.20,0.167,0.125,0.10,0.067\}$. It does therefore seem that the charge response is only very weakly affected by the formation of incommensurate spin correlations in this regime.

\begin{figure*}
\centering
\includegraphics[width=0.99\textwidth]{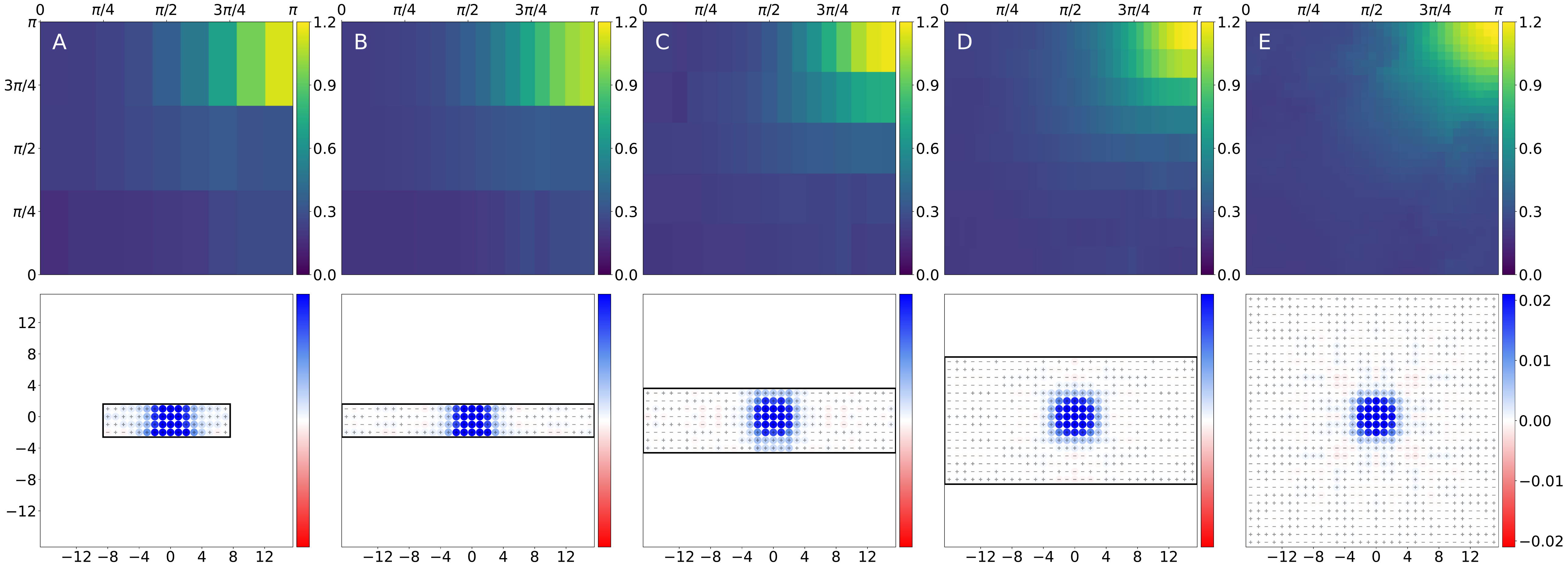}
\caption{Spin susceptibility in momentum space $\chispinq{\mathbf{q}}$ (top)
and staggered spin susceptibility in real-space $\chispinstag(\mathbf{r})$ (bottom) evaluated at $T = 0.22$ for $U=5$ and $n=0.875$ for different lattice geometries (with periodic boundary conditions) of sizes: $16\times4$~(A), $32\times4$~(B), $32\times8$~(C), $64\times16$~(D) and $64\times64$~(E). }
\label{Fig:system_size_rectangular_highT}
\end{figure*}

\begin{figure}
\centering
\includegraphics[width=0.42\textwidth]{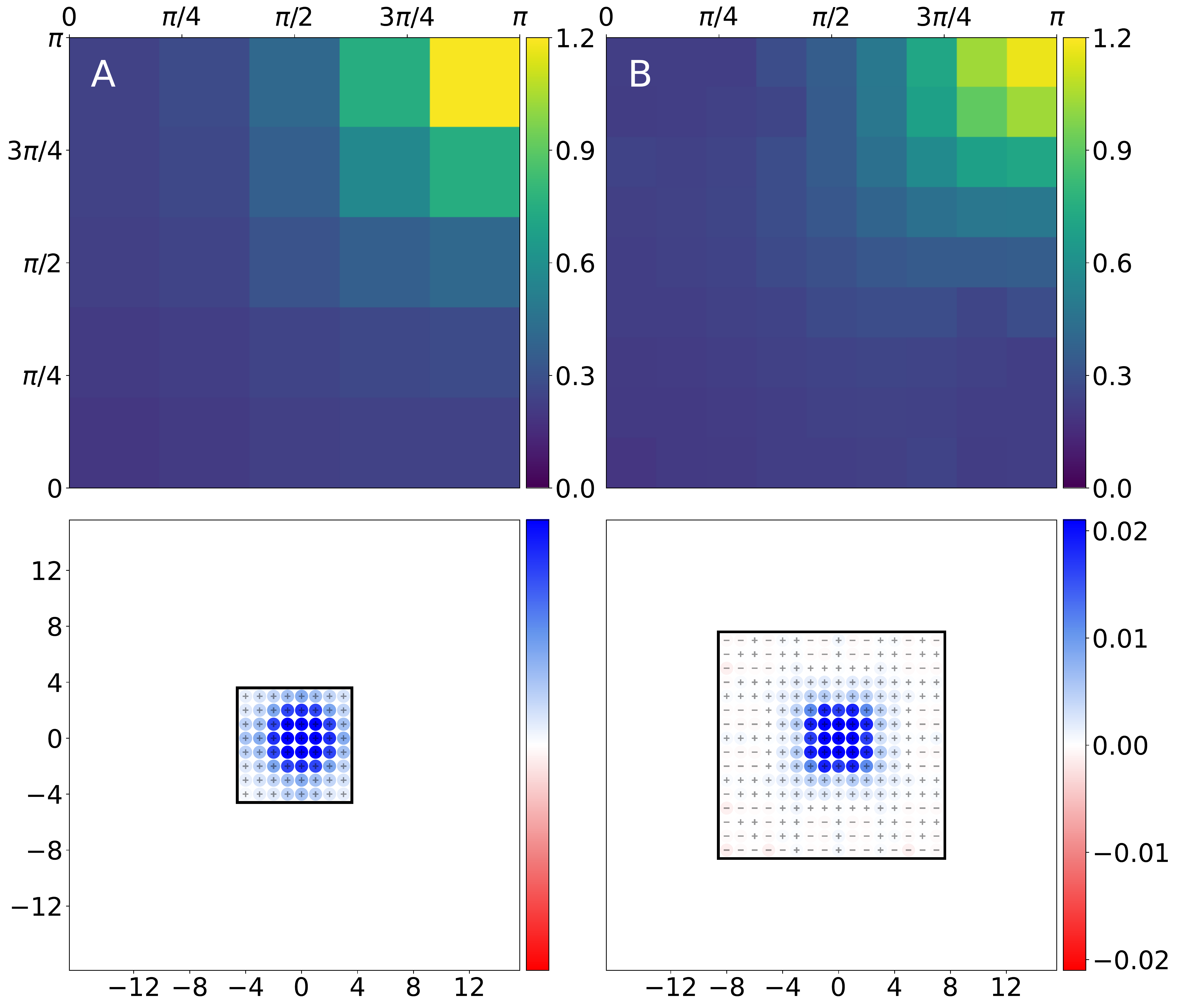}
\caption{Spin susceptibility in momentum space $\chispinq{\mathbf{q}}$ (top)
and staggered spin susceptibility in real-space $\chispinstag(\mathbf{r})$ (bottom) evaluated at $T = 0.22$ for $U=5$ and $n=0.875$ for two lattice geometries (with periodic boundary conditions) of sizes: $8\times8$~(A) and $16\times16$~(B).}
\label{Fig:system_size_square_highT}
\end{figure}

\section{System size dependence of the spin susceptibility at higher temperatures}
\label{app:system_size}

In Fig.~\ref{Fig:system_size_rectangular_highT} and Fig.~\ref{Fig:system_size_square_highT} we perform an additional study of the dependence of the spin susceptibility $\chi_{sp}$ in real and reciprocal space on variations in system size and lattice geometry (rectangular vs. square). We have picked the same density $n=0.875$ as in Figs.~\ref{Fig:system_size_rectangular},~\ref{Fig:system_size_square} of Sec.~\ref{sec:system_size}, however, we have chosen to investigate here the higher temperature $T=0.22$ (in accordance with Ref.~\cite{huang2018stripe}) and a somewhat higher interaction $U=5$ as it is closer to the maximum in correlation length as a function of interaction strength. We observe that all geometries manage to reproduce the correct value and position of the maximum in reciprocal space, which is centered around $(\pi,\pi)$, corresponding to commensurate correlations. This is consistent with the fact that lattice-size and boundary effect play an increasing role as correlations increase, which in what we universally observe when lowering the temperature.

\begin{figure*}
\centering
\includegraphics[width=0.99\textwidth]{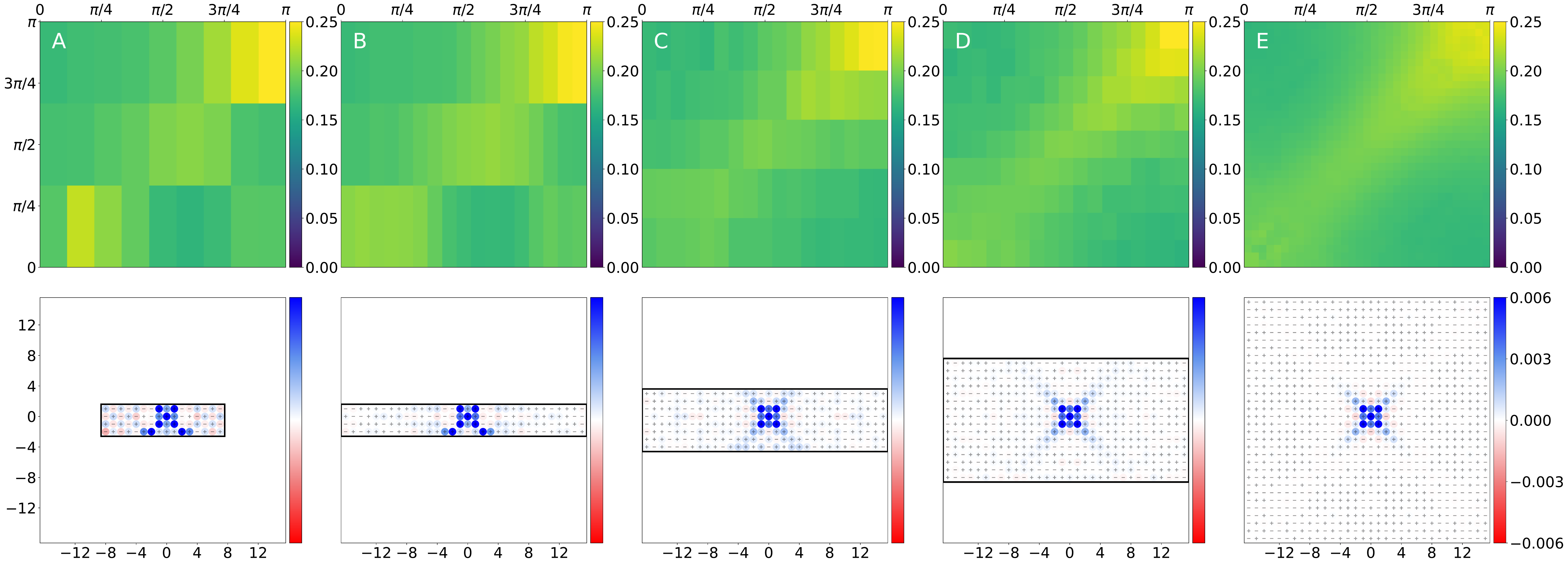}
\caption{Charge susceptibility in momentum space $\chi_\mathrm{ch}(\mathbf{q})$ (top)
and in real-space $\chi_\mathrm{ch}(\mathbf{r})$ (bottom) evaluated at $T = 0.1$ for $U=4$ and $n=0.875$ for different lattice geometries (with periodic boundary conditions) of sizes: $16\times4$~(A), $32\times4$~(B), $32\times8$~(C), $32\times16$~(D) and $64\times64$~(E). }
\label{Fig:system_size_rectangular_charge}
\end{figure*}

\begin{figure}
\centering
\includegraphics[width=0.42\textwidth]{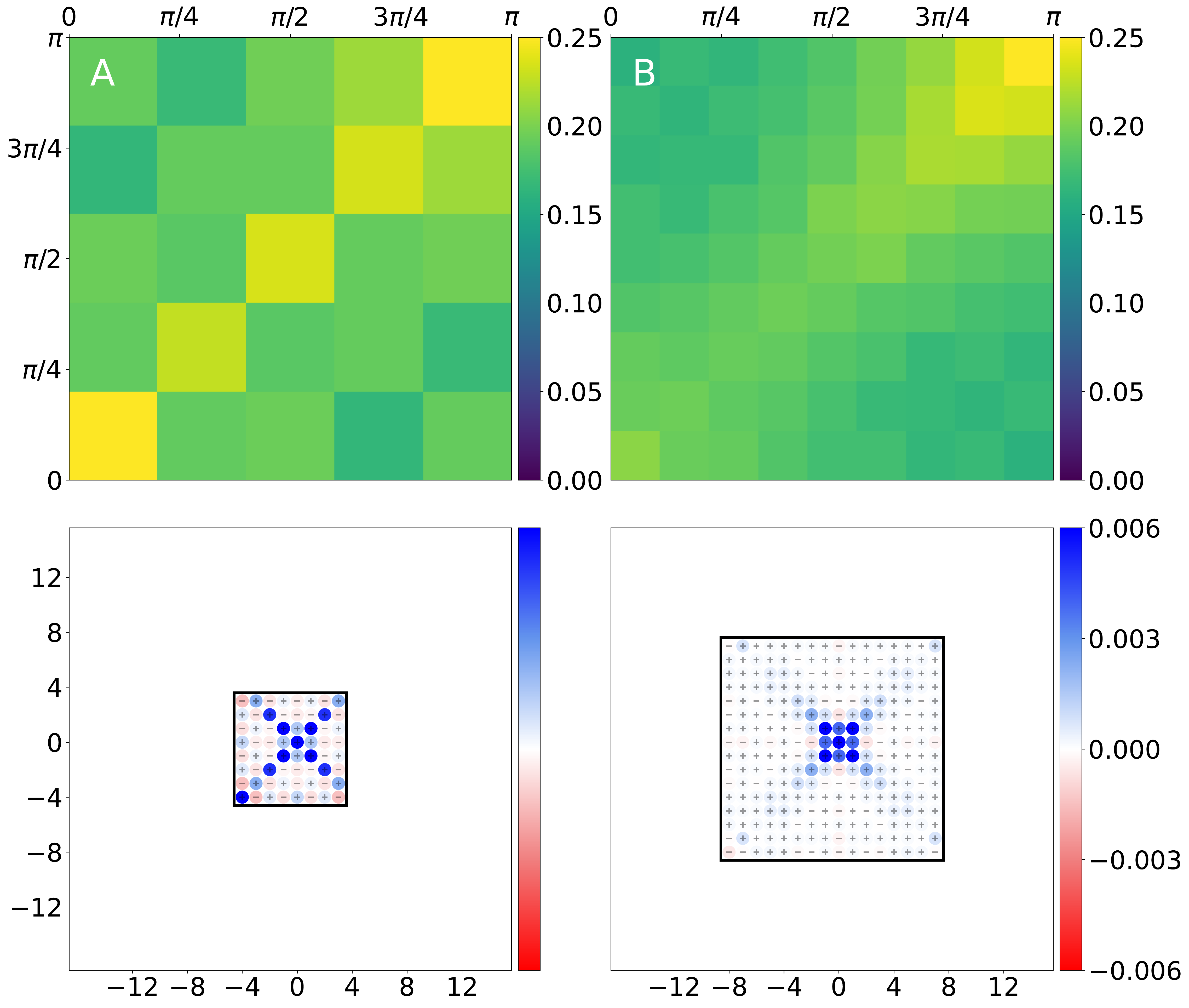}
\caption{Charge susceptibility in momentum space $\chi_\mathrm{ch}(\mathbf{q})$ (top)
and real-space $\chi_\mathrm{ch}(\mathbf{r})$ (bottom) evaluated at $T = 0.1$ for $U=4$ and $n=0.875$ for two lattice geometries (with periodic boundary conditions) of sizes: $8\times8$~(A) and $16\times16$~(B).}
\label{Fig:system_size_square_charge}
\end{figure}

In Fig.~\ref{Fig:system_size_rectangular_charge} and Fig.~\ref{Fig:system_size_square_charge} we equally analyse the system size effects of the charge susceptibility $\chi_{ch}$, at density $n=0.875$, temperature $T=0.1$ and interaction strength $U=4$. We observe that the overall real space picture is qualitatively similar for most lattices, even though the small geometries $16\times4$ and $8\times8$ show stronger non-local correlations as well as an enhanced peak around wave-vector $(0,0)$ as compared to the result on the large $64\times64$ lattice.

\section{Chemical-potential insertions as double expansions}
\label{app:cdet}

\subsection{Connected Determinant Diagrammatic Monte Carlo}
In this section we briefly reproduce the formalism of Ref.~\cite{cdet} in a form that is most convenient for its generalization. With the notation of Sec.~\ref{sec:formalism}, for any quantity $\mathcal{O}$ and for any real $\alpha$, we define
\begin{equation}
\mathcal{O}(\mu_0+\alpha\,U,U) = \sum_{k=0}^\infty U^k\,\mathcal{O}_{\alpha;k},
\end{equation}
where hereafter a dependence on the chemical potential at $U=0$, by $\mu_0$, is implied in all expressions.
It is possible to express $\mathcal{O}_{\alpha;k}$ as the integral of the sum of all connected Feynman diagrams with
$\{(\mathbf{r}_1,\tau_1),\dots,(\mathbf{r}_k,\tau_k)\}$ as space-time positions for the vertices
\begin{equation}
    \mathcal{O}_{\alpha;k}=
    \frac{1}{k!}\int_{\tau_1,\dots,\tau_k}\sum_{\mathbf{r}_1,\dots,\mathbf{r}_k} \mathcal{O}_{\alpha}(
    V
    ),
    \label{eq:coefficients-integral-cdet}
\end{equation}
where $V=\{(\mathbf{r}_1,\tau_1),\dots,(\mathbf{r}_k,\tau_k)\}$,
$\mathbf{r}_l$ is a lattice site and $\mathcal{\tau}_l\in\left[0,\frac{1}{T}\right]$ the imaginary-time. The integrand $\mathcal{O}_{\alpha}(\{(\mathbf{r}_1,\tau_1),\dots,(\mathbf{r}_k,\tau_k)\})$ can be computed numerically exactly for any quantity with exponentially scaling computational cost by using the CDet technique. As an example, we present here the computation of the density, $\mathcal{O}=n$. The CDet algorithm allows to obtain the sum of all connected Feynman diagrams, $n_{\alpha}(V)$, by removing disconnected diagrams from the sum of all, connected and disconnected, Feynman diagrams, which we denote here by $a_{\alpha}(V)$, and which has the advantage to be easily expressed in form of determinants thanks to the Wick's theorem. Indeed, one has the explicit form:
\begin{equation}
    a_{\alpha}(V)=\left(-1\right)^{|V|}\,\text{det} \,\mathbb{A}_{\alpha}(V)\;\text{det}\,\mathbb{Z}_{\alpha}(V),  \label{eq:a-cdet}
\end{equation}
where $V\equiv\{(\mathbf{r}_1,\tau_1),\dots,(\mathbf{r}_k,\tau_k)\}$, $|V|$ is the cardinality of the set $V$ corresponding to the expansion order, $\mathbb{A}_{\alpha}(V)$ is a $\left(|V|+1\right)\times \left(|V|+1\right)$ matrix given by:
\begin{equation}
    \left(\mathbb{A}_{\alpha}(V)\right)_{uv}=G_{}^{(0)}((\mathbf{r}_u,\tau_u),(\mathbf{r}_v,\tau_v))-\alpha \,\delta_{u,v} \,\bar{\delta}_{u,|V|+1},        \label{eq:a-det-cdet}
\end{equation}
where $\bar{\delta}_{u,v}=1-\delta_{u,v}$,
$\mathbb{Z}_{\alpha}(V)$ is a $|V|\times|V|$ matrix given by $ \left(\mathbb{Z}_{\alpha}\right)_{u,v}= \left(\mathbb{A}_{\alpha}\right)_{u,v}.$
We further set $\mathbf{r}_{|V|+1}\equiv\mathbf{0}$, $\tau_{|V|+1}\equiv0$, and the non-interacting Green's function is defined by:
\begin{equation}
    G_{}^{(0)}((\mathbf{r},\tau),(\mathbf{r}',\tau'))\equiv-\langle \operatorname{T}[c_\uparrow(\mathbf{r},\tau)
\,c_\uparrow^\dagger(\mathbf{r}',\tau')]\rangle_{\mu=\mu_0,\,U=0},
\end{equation}
where $\operatorname{T}$ is the time ordering operator and the average is computed at $\mu=\mu_0$ and $U=0$. Then, $n_{\alpha}(V)$ is obtained from the recursive elimination of disconnected diagrams from the sum of all diagrams:
\begin{equation}
    n_{\alpha}(V)
    =a_{\alpha}(V)-\sum_{S\subsetneq V} n_{\alpha}(S)\;z_{\alpha}(V\setminus S)        \label{eq:recursive-cdet}
\end{equation}
where we have introduced
\begin{equation}
    z_{\alpha}(V) \equiv\left(-1\right)^{|V|} \,\left(\det\,\mathbb{Z}_{\alpha}(V)\right)^2.
    \label{eq:z-cdet}
\end{equation}

The determinants of the matrices $\mathbb{A}_{\alpha}(S)$ and $\mathbb{Z}_{\alpha}(S)$ for all subsets $S\subseteq V$, needed for $a_{\alpha}(S)$ and $z_{\alpha}(S)$, can be computed in a computational time proportional to $O(2^k)$, where $k\equiv|V|$, by using a fast principal minor algorithm~\cite{griffin2006principal, vsimkovic2021fast}. This leaves the recursive step, Eq.~\eqref{eq:recursive-cdet}, as the main computational bottleneck with a complexity of $O(3^k)$ (or, alternatively, $O(k^2\, 2^k)$ using fast subset convolutions~\cite{koivisto}).

For completeness, we give the explicit expressions for the spin and charge susceptibilities at spin balance in terms of the $F^{\sigma\sigma'}$ functions:
\begin{equation}
\begin{split}
    &\chicharge(\mathbf{r})=2\int_0^{1/T}d\tau\left(
    F^{\uparrow\uparrow}(\mathbf{r},\tau)+
F^{\uparrow\downarrow}(\mathbf{r},\tau)
    \right),\\
    &\chispin(\mathbf{r})=\frac{1}{2}\int_0^{1/T}d\tau\left(
    F^{\uparrow\uparrow}(\mathbf{r},\tau)-
F^{\uparrow\downarrow}(\mathbf{r},\tau)
    \right),
    \end{split}
\end{equation}
where
\begin{equation}
      F^{\sigma\sigma'}(\mathbf{r},\tau)= \langle\delta\hat{n}_\sigma(\mathbf{r},\tau)\,\delta\hat{n}_{\sigma'}(\mathbf{0},0)\rangle.
\end{equation}
The diagrammatic expansions for $F^{\uparrow\uparrow}$ and $F^{\uparrow\downarrow}$ can be found in Fig.~\ref{Fig:feynman-diagrams-alpha}.
We write the equivalent of Eq.~\eqref{eq:coefficients-integral-cdet} for $F^{\sigma\sigma'}$:
\begin{equation}
    F^{\sigma\sigma'}_{\alpha;k}=
    \frac{1}{k!}\int_{\tau_1,\dots,\tau_k}\sum_{\mathbf{r}_1,\dots,\mathbf{r}_k} F^{\sigma\sigma'}_{\alpha}(
    V
    ).
    \label{eq:coefficients-integral-cdet-F}
\end{equation}
In order to obtain $F^{\sigma\sigma'}_{\alpha}(V)$, analogously to what was done before, we introduce
\begin{equation}
\begin{split}
    &A^{\uparrow\uparrow}_{\alpha}(V)=\left(-1\right)^{|V|}\,\text{det} \,\mathbb{B}_{\alpha}(V)\,
\text{det} \,\mathbb{Z}_{\alpha}(V),\\
    &A^{\uparrow\downarrow}_{\alpha}(V)=\left(-1\right)^{|V|}\,
    \text{det} \,\tilde{\mathbb{A}}_{\alpha}(V)
\,\text{det} \,\mathbb{A}_{\alpha}(V),
\end{split}
    \label{eq:A-cdet}
\end{equation}
where $\mathbb{B}_{\alpha}$ is a $(|V|+2)\times(|V|+2)$ matrix defined by
\begin{equation}
\begin{split}
 \left(\mathbb{B}_{\alpha}(V)\right)_{uv}&=G_{}^{(0)}((\mathbf{r}_u,\tau_u),(\mathbf{r}_v,\tau_v))\\
 &-\alpha \,\delta_{u,v} \,\bar{\delta}_{u,|V|+1} \bar{\delta}_{u,|V|+2},
 \end{split}
 \label{eq:b-det-cdet}  \end{equation}
where $\mathbf{r}_{|V|+1}=\mathbf{r}$, $\tau_{|V|+1}=\tau$, $\mathbf{r}_{|V|+2}=\mathbf{0}$, $\tau_{|V|+2}=0$, and $\tilde{\mathbb{A}}_{\alpha}$ is a $(|V|+1)\times(|V|+1)$ matrix such that $(\tilde{\mathbb{A}})_{uv} = (\mathbb{B})_{uv}$. We also define:
\begin{equation}
    \tilde{a}_{\alpha}(V)=\left(-1\right)^{|V|}\,\text{det} \,\tilde{\mathbb{A}}_{\alpha}(V)\;\text{det}\,\mathbb{Z}_{\alpha}(V). \label{eq:a-tilde-cdet}
\end{equation}
We can then obtain $F_{\sigma\sigma'}$ from the recursive relation:
\begin{equation}
\begin{split}
    F^{\sigma\sigma'}_{\alpha}(V)&=A^{\sigma\sigma'}_{\alpha}(V)\\
    &-\sum_{S\subseteq V} n_{\alpha}(S)\,\tilde{a}_{\alpha}(V\setminus S)\\
    &-\sum_{S\subsetneq V}F^{\sigma\sigma'}_{\alpha}(S)\,z_{\alpha}(V\setminus S),
\end{split}
\end{equation}
which has the diagrammatic interpretation of the elimination from the sum of all, connected and disconnected, symmetrized Feynman diagrams for $\langle \hat{n}_{\sigma}(\mathbf{r},\tau)\,\hat{n}_{\sigma'}(\mathbf{0},0)\rangle$, $A^{\sigma\sigma'}_{\alpha}(V)$, of two classes of diagrams: the non-necessarily connected diagrams contributing to $\langle\hat{n}_\sigma(\mathbf{r},\tau)\rangle\langle\hat{n}_{\sigma'}(\mathbf{0},0)\rangle$ (second line in the previous equation), and the remaining disconnected diagrams (third line in the previous equation). The equivalent CDet recursion formula for single Hartree-type expansions of the spin and charge susceptibilities has been first introduced in Ref.~\cite{kim_cdet}.

\subsection{Evaluating arbitrary chemical-potential insertions efficiently with CDet}
Equations \eqref{eq:a-cdet}, \eqref{eq:a-det-cdet} and \eqref{eq:z-cdet} show that $a_{\alpha}(V)$ and $z_{\alpha}(V)$ are polynomials of $\alpha$ of degree $2\,|V|$
\begin{equation}
\begin{split}
&a_{\alpha}(V)=\sum_{j=0}^{2|V|}\alpha^j\,a_{j}(V),\\
&z_{\alpha}(V)=\sum_{j=0}^{2|V|}\alpha^j\,z_{j}(V),
\end{split}
\end{equation}
and Eq.~\eqref{eq:recursive-cdet} shows that $n_{\alpha}(V)$ is a polynomial of $\alpha$ of degree, at most $2\,|V|$, and from the Feynman-diagram representation it is easy to understand that the actual degree is only $|V|$:
\begin{equation}
    n_{\alpha}(V)=\sum_{j=0}^{|V|}\alpha^j\, n_{j}(V).
\end{equation}
The same reasoning applies to any quantity $\mathcal{O}$: $\mathcal{O}_{\alpha}(V)=\sum_{j=0}^{|V|}\alpha^j\, \mathcal{O}_{j}(V)$, where
$\mathcal{O}_{j}(V)$ is the $j$-th chemical potential insertion
in the sum of all connected Feynman diagrams for fixed space-time vertex positions $V$, from which we can determine the contribution to the physical quantity $\mathcal{O}$ after integration over space-time coordinates. In the following, we limit our discussion to $\mathcal{O}=n$.

\begin{figure}
\centering
\includegraphics[width=0.4\textwidth]{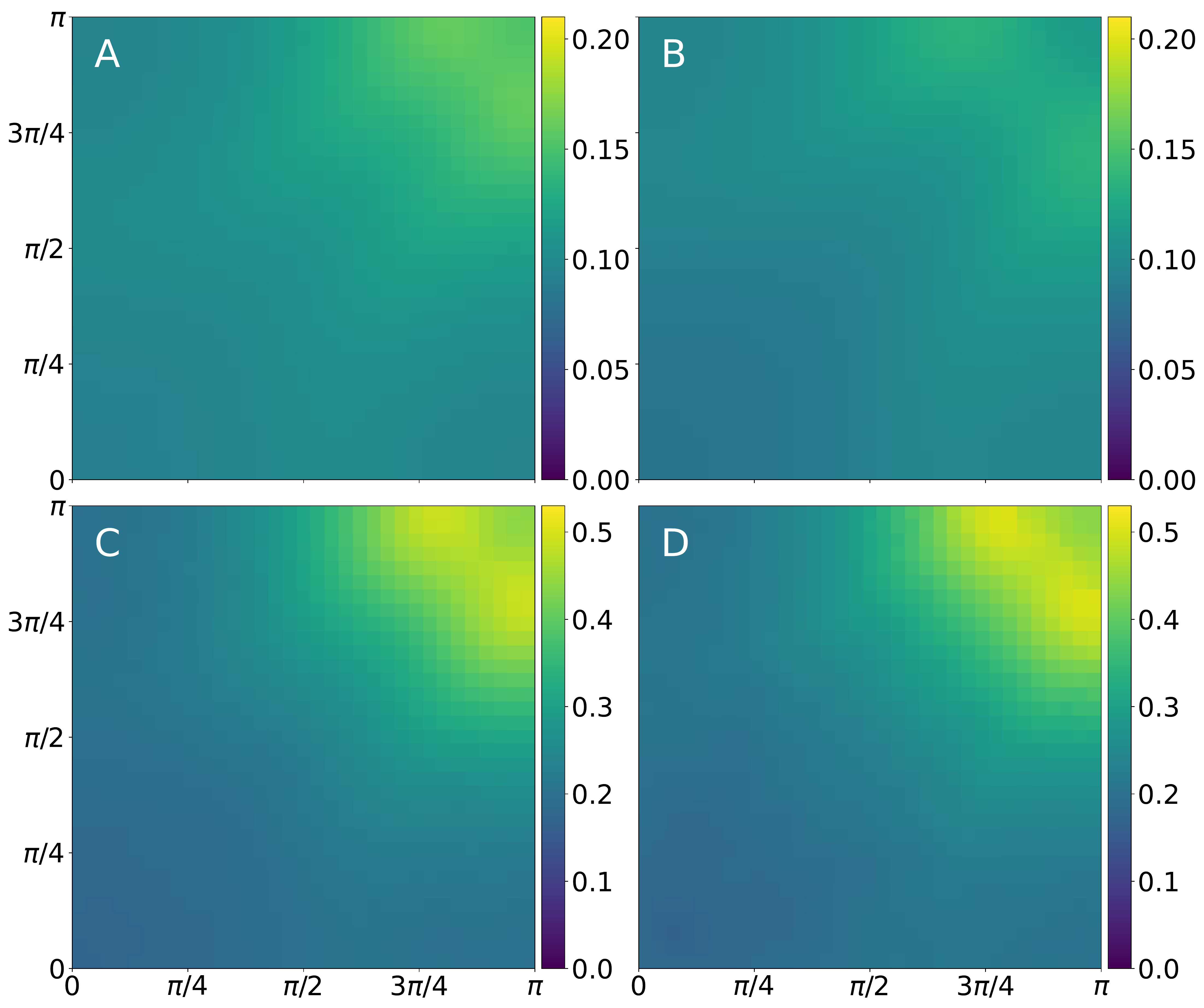}
\caption{Spin susceptibility $\chispinq{q}$ for $n=0.775$, $T=0.2$, $U=4.35$ (bottom row), computed from the fixed-density series (left) as well as from a Hartree-shifted series (right). The $U=0$ starting points of the perturbation series
are shown in the top row and correspond to densities $n_0=0.775$ (left) and $n_0=0.7$ (right).}
\label{Fig:comparison_fixed_n_mu}
\end{figure}

Our strategy to compute $n_{j}(V)$, for $j\le |V|$, is the following: we first determine $a_{j}(S)$ and $z_{j}(S)$ for all $S\subseteq V$ and $j\le |V|$; then, we apply Eq.~\eqref{eq:recursive-cdet} seen as a {\it polynomial} equation to recursively obtain $n_{j}(V)$. We use the following identities:
\begin{equation}
\begin{split}
    &\text{det}\,\mathbb{A}_{\alpha}(S)=\sum_{j=0}^{|S|}\alpha^j\,\sum_{S'\subseteq S:\,|S'|=|S|-j}\text{det}\, \mathbb{A}_{0}(S'),\\
    &\text{det}\,\mathbb{Z}_{\alpha}(S)=\sum_{j=0}^{|S|}\alpha^j\,\sum_{S'\subseteq S:\,|S'|=|S|-j}\text{det}\, \mathbb{Z}_{0}(S'),
\end{split}
\label{eq:det-alpha-subset}
\end{equation}
which can be evaluated for all $S\subseteq V$ in $O(3^{|V|})$ computational steps (or in $O(|V|^2\,2^{|V|})$ by using ranked zeta transforms~\cite{koivisto}). Using Eq.~\eqref{eq:a-cdet}, \eqref{eq:z-cdet} and~\eqref{eq:det-alpha-subset} we can build the $a_{\alpha}(S)$,  $z_{\alpha}(S)$ polynomials, which can then be used to obtain the polynomial $n_{\alpha}(V)$, and therefore $n_{j}(V)$, from Eq.~\eqref{eq:recursive-cdet}, with a computational cost $O(|V|^2\,3^{|V|})$. An analogous strategy applies to the computation of $\mathcal{O}_{j}(S)$, and the final contribution to the $\mathcal{O}$ quantity can be evaluated from Eq.~\eqref{eq:coefficients-integral-cdet}
\begin{equation}
    \mathcal{O}_{\alpha;k}=
    \frac{1}{k!}\sum_{j=0}^{k}\alpha^j\int_{\tau_1,\dots,\tau_k}\sum_{\mathbf{r}_1,\dots,\mathbf{r}_k} \mathcal{O}_{j}(V),
    \label{eq:coefficients-integral-cdet-alpha}
\end{equation}
where $V=\{(\mathbf{r}_1,\tau_1),\dots,(\mathbf{r}_k,\tau_k)\}$.

As briefly mentioned in the main text, one of the features of the algorithmic advancement we discuss here is that it allows to crosscheck the results obtained by different choices of $\alpha$. This freedom has been proven to be extremely useful~\cite{rubtsov2005continuous, wu_controlling}, even if in these previous works it was impossible to use the full functional freedom of $\alpha(U)$ as a function of $U$, as there was no technical means to achieve this goal efficiently. In this work, we have limited ourselves to comparing the Hartree-shift choice, $\alpha=n_0/2$, and the fixed-density choice, i.e. $\alpha(U)$ such that the density does not change as a function of $U$. See Fig.~\ref{Fig:comparison_fixed_n_mu} for a representative comparison of the two set of series for different wave-vectors of the upper quarter of the Brillouin zone. These have two fairly different starting points, yet are converging to the same results within error bars.

It is also possible to generalize this formalism to arbitrary symmetry-breaking field insertions. A constant symmetry-breaking field shift has already been shown to be an efficient way to build a convergent perturbative series in the superfluid phase~\cite{spada2021highorder}. We further remark that a similar technique to the one we introduce in this work can be applied to a more general shift in the single-particle propagator, which would allow even more flexibility in the control of the convergence of the series. For completeness we note that alternative exponentially scaling algorithms which alter perturbative expansions whilst still grouping Feynman diagrams into determinants have been introduced in Refs.~\cite{rdet, rpadet, kim2021homotopic}.

\newpage

\bibliographystyle{ieeetr}
\bibliography{main_biblio}

\end{document}